\documentclass[pre,twocolumn,amsmath]{revtex4-1}

\usepackage{graphicx}

\usepackage{amssymb,amsfonts,amsmath}

\def \>{\rangle} 
\def \<{\langle} 
 
\def\be{\begin{equation}} 
\def\ee{\end{equation}} 
\def\longrightharpoonup{\relbar\joinrel\rightharpoonup}
\def\longleftharpoondown{\leftharpoondown\joinrel\relbar}

\def\longrightleftharpoons{
  \mathop{
    \vcenter{
      \hbox{
      \ooalign{
        \raise1pt\hbox{$\longrightharpoonup\joinrel$}\crcr
	  \lower1pt\hbox{$\longleftharpoondown\joinrel$}
	  }
      }
    }
  }
}

\newcommand \bea {\begin{eqnarray}} 
\newcommand \eea {\end{eqnarray}}

\begin{document}

\title{An analytically tractable model for community ecology with many species}

\author{Benjamin Dickens}
\affiliation{Dept. of Physics, Boston University, Boston, MA 02215}

\author{Charles K. Fisher}
\email{ Current address: Pfizer, Cambridge, MA }
\affiliation{Dept. of Physics, Boston University, Boston, MA 02215}

\author{Pankaj Mehta}
\email{pankajm@bu.edu}
\affiliation{Dept. of Physics, Boston University, Boston, MA 02215}

\begin{abstract}
A fundamental problem in community ecology is to understand how ecological processes such as  selection, drift, and immigration give rise to observed patterns in species composition and diversity. Here, we present a simple, analytically tractable, presence-absence (PA) model for community assembly and use it to ask how ecological traits such as  the strength of competition, the amount of diversity, and demographic and environmental stochasticity affect species composition in a community.  In the PA model, species are treated as stochastic binary variables that can either be present or absent in a community: species can immigrate into the community from a regional species pool and can go extinct due to competition and stochasticity. Despite its simplicity, the PA model reproduces the qualitative features of  more complicated models of community assembly. In agreement with recent work on large, competitive Lotka-Volterra systems, the PA model exhibits distinct ecological behaviors organized around a special (``critical") point corresponding to Hubbell's neutral theory of biodiversity. These results suggest that the concepts of ecological ``phases'' and phase diagrams can provide a powerful framework for thinking about community ecology and that the PA model captures the essential ecological dynamics of community assembly.
 \end{abstract}

\maketitle

A central goal of community ecology is to understand the tremendous biodiversity present in naturally occurring communities. The observed patterns of species composition and diversity stem from the interaction of a number of ecological processes. Traditional models of community assembly (referred to as `niche' models) emphasize the important role played by competition and ecological selection in shaping community structure \cite{tilman_resource_1982, hardin_competitive_1960, chesson_macarthurs_1990,macarthur1970species,macarthur_limiting_1967}. However, due to the introduction of the neutral theory of biodiversity, the past fifteen years have seen a renewed interest in the role of drift, or stochasticity, in shaping community assembly  \cite{hubbell_unified_2001, volkov_neutral_2003, rosindell_unified_2011,rosindell_case_2012}. In the neutral theory, all species have identical birth and death rates so that all variation in species abundances is due entirely to random processes. A complete theory of community assembly must take into account other ecological processes, such as immigration and speciation, in addition to selection and drift \cite{macarthur1963equilibrium, vellend_conceptual_2010}. This has led to a renewed interest in using methods from statistical physics to understand the basic principles governing community assembly  \cite{ rulands2013global, fisher2014transition,kussell2014non, kessler2015generalized, azaele2015statistical, kalyuzhny2014niche}.

One common approach for modeling community assembly in complex communities is to consider generalized Lotka-Volterra models (LVMs) \cite{may_will_1972, fisher2014transition, kessler2015generalized}. In generalized LVMs, ecological dynamics are modeled using a system of non-linear differential equations for species abundances. Each species is characterized by a carrying capacity -- i.e., its maximal population size in absence of other species. Species interactions are modeled using a matrix of  ``interaction coefficients.'' In general, it is extremely difficult to precisely measure these interaction coefficients \cite{fisher2014identifying}. However, for ecosystems with many species, we can overcome this difficulty by considering a ``typical ecosystem'' for which species interaction matrices are drawn from a random matrix ensemble \cite{may_will_1972}. 

Historically, LVMs emphasized the role of ecological selection and resource availability. For this reason, LVMs were traditionally analyzed as deterministic ordinary differential equations (ODEs). However, several recent studies have moved beyond deterministic ODE models to incorporate the effects of immigration and stochasticity on ecological dynamics \cite{fisher2014transition, kessler2015generalized}. These recent studies on typical ecosystems have demonstrated that communities can exhibit distinct ecological ``phases" (i.e., regimes with qualitatively different species abundance patterns) as ecological parameters such as immigration rates and the strength and heterogeneity of competition are varied. For example, by numerically simulating stochastic differential equation-based implementation of a generalized LVM, \cite{fisher2014transition} showed that communities can exhibit a sharp transition between a selection-dominated regime dominated by a single stable equilibrium and a drift-dominated regime where species abundances are uncorrelated and the ecological dynamics is well approximated by neutral models. The selection-dominated regime is favored in communities with large population sizes and relatively constant environments, whereas the neutral phase is favored in communities with small population sizes and fluctuating environments. Similarly, \cite{ kessler2015generalized} used a stochastic LVM to analyze a local community of competing species with weak immigration from a static regional pool and identified four distinct ecological phases organized around a ``critical point'' corresponding to Hubbell's neutral model \cite{hubbell_unified_2001}.

Although LVMs are among the standard tools of theoretical ecology, they are difficult to analyze with analytic techniques -- especially in the stochastic setting. For this reason,  \cite{fisher2014transition} introduced an immigration-extinction process (referred to as the presence-absence (PA) model) for community assembly that attempts to capture the essential ecology of LVMs using a simpler model. In the PA model, species are treated as stochastic binary variables that are either present or absent in a local community. A species may become extinct (i.e., absent) in the local community due to competitive exclusion and stochasticity, but it can reappear in the community by immigrating from a regional species pool. In contrast to LVMs, the PA model is amenable to analytical arguments using techniques from statistical physics related to the study of disordered spin systems. For example, the aforementioned sharp transition between the selection-dominated regime and the drift-dominated regime seen in generalized LVMs corresponds to the analogue of the ``freezing transition'' in the PA model \cite{fisher2014transition}.

In this work, we address the extent to which the PA model reproduces the qualitative behaviors and ecological regimes found in more complicated LVMs. To address this question, we use the PA model to analyze a local community of competing species with weak immigration from a static regional pool and compare it to the ecological dynamics seen in numerical simulations of the generalized LVMs \cite{kessler2015generalized}. We numerically simulate the PA model and construct phase diagrams for species abundances to see how ecological processes such as  selection, drift, and immigration affect species abundance patterns. We supplement these results with analytic arguments. We then compare and contrast the phase diagrams obtained from the PA model and LVMs. Finally, we discuss the implications of our results for modeling complex ecological communities.

\section{The Presence-Absence (PA) Model}

\begin{figure}[t!]
\includegraphics[width=0.5\textwidth]{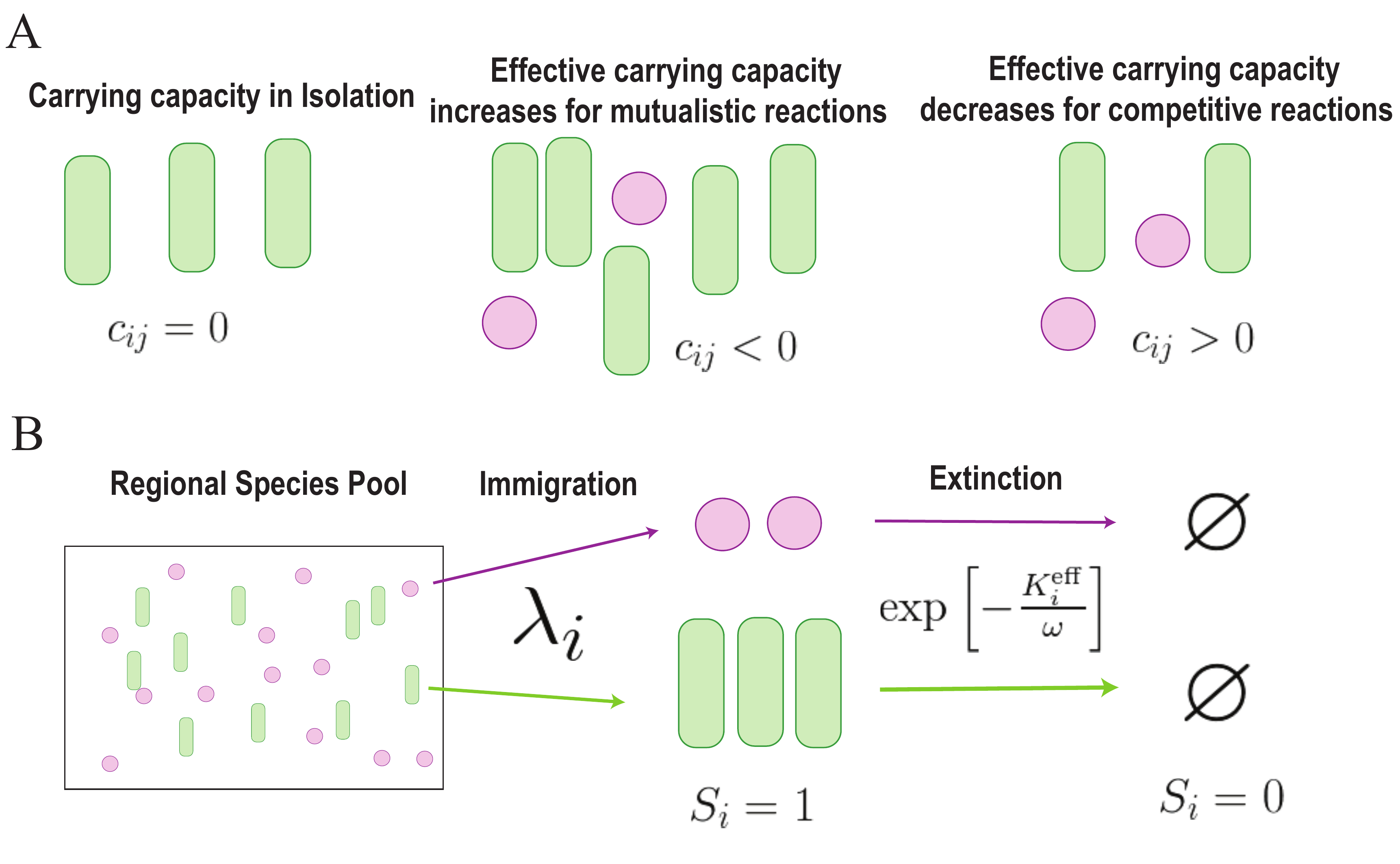}
\caption{{\bf The presence-absence model.}  (A)  Species $i$ has a carrying capacity, $K_i$, in the absence of other species. Species $i$'s interaction with species $j$ is characterized by the interaction coefficient, $c_{ij}$. Depending on whether they are mutualistic or competitive, the interactions can increase or decrease the effective carrying capacity of species $i$. (B)  Species $i$ can be present in the community, $s_i=1$, or absent from the community, $s_i = 0$. Species can immigrate from a regional community pool and go extinct in the community proportional with a rate equal to the exponential of the ratio of effective carrying capacity to the stochasticity parameter, $\omega$. }
\label{schematic}
\end{figure}

\subsection{Ecological motivation}

The goal of introducing the PA model is to capture the essential ecological features present in LVMs in a simple, analytically tractable model. For the sake of tractability, the PA model ignores species abundances and instead focuses on a simpler question: is a species present or absent in the community? The basic idea behind the definition of the PA model is, roughly speaking, that the propensity of a species to be present or absent is determined by a quantity called its ``effective carrying capacity." The effective carrying capacity of a species $i$, which generally depends on both environmental factors as well as the abundances of other species, sets the maximum possible abundance of a species $i$ in the presence of the others. Therefore, species $i$ may persist if its effective carrying capacity is positive, whereas it will go extinct if its effective carrying capacity is negative. 

To gain intuition about the role of effective carrying capacities in the definition of the PA model, it is helpful to recount the results on species invasion in Lotka-Volterra communities derived by MacArthur and Levins in their classic 1967 paper \cite{macarthur_limiting_1967}, wherein effective carrying capacities played a crucial role. Species abundances, $\vec{x}$, are modeled using a system of  ODEs of the form $ d x_i / dt = \lambda_i + x_i f_i(\vec{x})$, with $\lambda_i$ the rate of immigration and $f_i(\vec{x})$ the ecological fitness of species $i$, which is a function of the species abundances $\vec{x}$. In general, $f_i(\vec{x})$ may be a complicated function due to nonlinear functional responses or other phenomena. Regardless of the exact form of $f_i(\vec{x})$, the ecological fitness can always be linearized near an equilibrium point, $\vec{x}^{\,*}$, where the dynamics are approximately described by LVM equations. 

In the LVM, the ecological fitness, $f_i(\vec{x})= K_i - x_i -\sum_{j} c_{ij} x_j$, is a linear function of the carrying capacity, $K_i$, and interaction coefficients, $c_{ij}$, which measure how the presence of species $j$ affects the growth rate of species $i$. The interaction coefficients, $c_{ij}<0$, are negative when interactions with species $j$ benefit the growth of species $i$,  $c_{ij}>0$ when species $j$ competes with species $i$, and $c_{ij}=0$ if species $i$ and $j$ do not interact (see Figure \ref{schematic}). We interpret $f_i(\vec{x})+x_i= K_i -\sum_{j} c_{ij} x_j$ as an effective carrying capacity, $K_i^{\mathrm{eff}}(\vec{x})$, for species $i$ and write $ d x_i / dt = \lambda_i + x_i(K_i^{\mathrm{eff}}(\vec{x})-x_i)$. In general, the effective carrying capacity is a function of the abundances of all the species in the community.

MacArthur and Levins \cite{macarthur_limiting_1967} used the idea of an effective carrying capacity to ask whether a new species $i$ could invade a community with species abundances $\vec{x}^{\,*}$. Using graphical stability arguments, they showed that species $i$ can invade successfully if its effective carrying capacity is positive ($K_i^{\mathrm{eff}}(\vec{x}^{\,*})= K_i -\sum_{j} c_{ij} x_j^* > 0$) but will be unsuccessful if its effective carrying capacity is negative ($K_i^{\mathrm{eff}}(\vec{x}^{\,*}) =K_i -\sum_{j} c_{ij} x_j^* < 0$). Therefore, the mean extinction time of a species in the local community depends strongly on the effective carrying capacity; the time to extinction is long for $K_i^{\mathrm{eff}}(\vec{x}^{\,*})>0$ and short for $K_i^{\mathrm{eff}}(\vec{x}^{\,*})<0$.  Based on these observations, we hypothesize that the extinction rate of a species depends exponentially on the effective carrying capacity. This assumption is used in the definition of the PA model.  

\subsection{Definition of the PA model}

The PA model describes the probability that various collections of species will be present (or absent) in a local ecological community, which we assume is attached to a large regional species pool containing $S$ species. We parametrize the presence (or absence) of species $i\in\left\{{1,...,S}\right\}$ by a binary random variable $s_i$, where $s_i=1$ if species $i$ is present and $s_i=0$ if it is absent. Therefore, the state of the ecosystem is described by the random vector $\vec{s}=(s_1,...,s_S)\in\left\{{0,1}\right\}^S$. We denote the probability distribution to observe a particular state $\vec{s}$ at time $t$ by $P_t(\vec{s})$. The probability distribution is governed by a differential equation called a master equation, which defines the dynamics of the PA model.

Prior to writing down the master equation, we specify two kinds of rates. First, there is the rate at which species $i$ immigrates into the local community from the regional pool, i.e. the rate at which $s_i = 0 \rightarrow 1$:
\begin{align*}
	R_i^I(\vec{s})=\lambda_i.
\end{align*}
There is also the rate of an extinction event $s_i=1\rightarrow 0$, $R_i^E(\vec{s})$, given by
\begin{align*}
	R_i^E(\vec{s})=\exp\left(-\frac{1}{\omega}K_i^{\mathrm{eff}}(\vec{s})\right),\\
	K_i^{\mathrm{eff}}(\vec{s}):=K_i-\sum_{j=1}^{j=S}K_jc_{ij}s_j.
\end{align*}
Here, $K_i^{\mathrm{eff}}(\vec{s})$ represents the effective carrying capacity of species $i$ given that the state of the ecosystem is $\vec{s}$. $K_i$ denotes the carrying capacity of species $i$ in the absence of other species, whereas $c_{ij}$ denotes an interaction coefficient describing how species $j$ influences the effective carrying capacity of species $i$ (with the convention that $c_{ii}=0$). The number $\omega$ parametrizes the impact of random noise on species extinction, and is thus called the ``noise strength." The units of time have been set so that the rate of extinction equals one in the limit that $\omega\rightarrow\infty$.

With these rates, the time evolution of $P_t(\vec{s})$ is given by the master equation:
\bea
	\frac{dP_t(\vec{s})}{dt} &=&\sum_{i=1}^{i=S}[(R_i^E(\vec{s}+\vec{e_i})P_t(\vec{s}+\vec{e_i})-R_i^I(\vec{s})P_t(\vec{s}))(1-s_i) \nonumber \\
	&+& (R_i^I(\vec{s}-\vec{e_i})P_t(\vec{s}-\vec{e_i})-R_i^E(\vec{s})P_t(\vec{s}))s_i],
\eea
where $\vec{e_i}$ denotes the vector whose $i$-th component is unity and all other components are zero. 

\subsection{Choosing carrying capacities and interaction coefficients}

The ecological dynamics of the PA model depend on the choice of carrying capacities and interaction coefficients. For an ecosystems with $S$ species, this involves specifying $S^2$ parameters. Deriving all of the parameters describing the dynamics of a real community from observations is a daunting task for ecosystems with many species ($S \gg 1$). However, it is possible to make progress by analyzing a ``typical'' ecosystem where the interaction coefficients and carrying capacities are drawn randomly from an appropriate probability distribution \cite{may_will_1972}. 

For simplicity, we restrict our analysis to purely competitive species interactions $c_{ij}>0$. We draw interaction coefficients independently for each pair $i\not=j$, from a gamma distribution with mean $\mu_c/S$ and variance $\sigma_c^2/S$:
\begin{align*}
	p_c(c_{ij})=\frac{1}{\theta_c^{k_c}\Gamma(k_c)}c_{ij}^{k_c-1}\exp\left(-\frac{c_{ij}}{\theta_c}\right),
\end{align*}
where $\Gamma$ denotes the Gamma function and	
\begin{align*}
	k_c:=\frac{\mu_c^2}{S\sigma_c^2},\;\;\;\; \theta_c:=\frac{\sigma_c^2}{\mu_c}.
\end{align*}
The $1/S$ scaling of the mean and variance of $c_{ij}$ is necessary to prevent pathological behaviors when $S$ becomes large. In physics terminology, this ensures a well-defined thermodynamic limit \cite{sherrington1975solvable}.

The carrying capacities are also drawn independently from a log-normal distribution with mean $\mu_K$ and variance $\sigma_K^2$:
\begin{align*}
	p_K(K_i)=\frac{1}{K_iz_K\sqrt{2\pi}}\exp\left(-\frac{1}{2z_K^2}[\mathrm{ln}(K_i)-l_K]^2\right),
\end{align*}
where
\begin{align*}
	l_K:=\mathrm{ln}\left(\frac{\mu_K^2}{\sqrt{\mu_K^2+\sigma_K^2}}\right),\;\;\;\;
		z_K:=\sqrt{\mathrm{ln}\left(1+\frac{\sigma_K^2}{\mu_K^2}\right)}.
\end{align*}

These choices of probability distributions ensure that both the interaction coefficients and the carrying capacities are strictly positive while simultaneously allowing for analytic calculations. Considering typical ecosystems for which $K_i$ and $c_{ij}$ are random variables circumvents the proliferation of free parameters by reducing the number of relevant parameters from $S^2$ to four: the means and variances of the interaction coefficients and carrying capacities $\mu_c, \sigma_c, \mu_K,$ and $\sigma_K$. 

\subsection{Relation to island biogeography}

The PA model describes the dynamics of a well-mixed, isolated community of competing species with weak immigration from a static regional pool. For this reason, the model is well-suited for discussions in the context of island biogeography. Island biogeography, the study of the species richness and ecological dynamics of isolated natural communities \cite{macarthur1963equilibrium, macarthur1967theory}, has played an important role in the development of theoretical ecology. For example, it was a precursor to Hubbell's neutral theory \cite{hubbell_unified_2001}. The success of the neutral theory of biodiversity and biogeography  \cite{rosindell_case_2012, hubbell_unified_2001} at explaining patterns in biodiversity has resulted in a vigorous debate on the processes underlying community assembly and, in particular, on the relative importance of selection and stochasticity in shaping ecological dynamics and species abundance patterns \cite{rosindell_case_2012, ricklefs_global_2012, mcgill_test_2003, ricklefs_unified_2006, dornelas_coral_2006,jeraldo2012quantification, tilman_niche_2004, volkov2009inferring, haegeman2011independent, chisholm_niche_2010, azaele2015statistical}. Overall, insular communities provide a tractable arena for studying the effects of selection and stochasticity while minimizing the effect of other ecological processes, such as complicated dispersal phenomena. 

The PA model allows one to study the roles of selection and stochasticity within the context of island biogeography. In particular, one may use the PA model to describe the dynamics of a local island community of competing species with weak immigration from a static regional pool, i.e a nearby mainland. This situation was recently analyzed using LVMs and found to exhibit distinct regimes of ecological dynamics and species abundances centered around a special critical point corresponding to Hubbell's neutral theory of biodiversity \cite{kessler2015generalized}. Inspired by earlier work showing that the PA model can reproduce the sharp transitions between a niche-like selection-dominated regime and a neutral-like drift-dominated regime \cite{fisher2014transition}, we numerically simulated the PA model to test whether or not it can reproduce the basic phenomenology seen in much more complicated LVMs. This is discussed in the next section.

\section{Numerical simulations}

\begin{figure*}[t]
	\centering
	\begin{tabular}{lll}
   		\includegraphics[scale=.28]{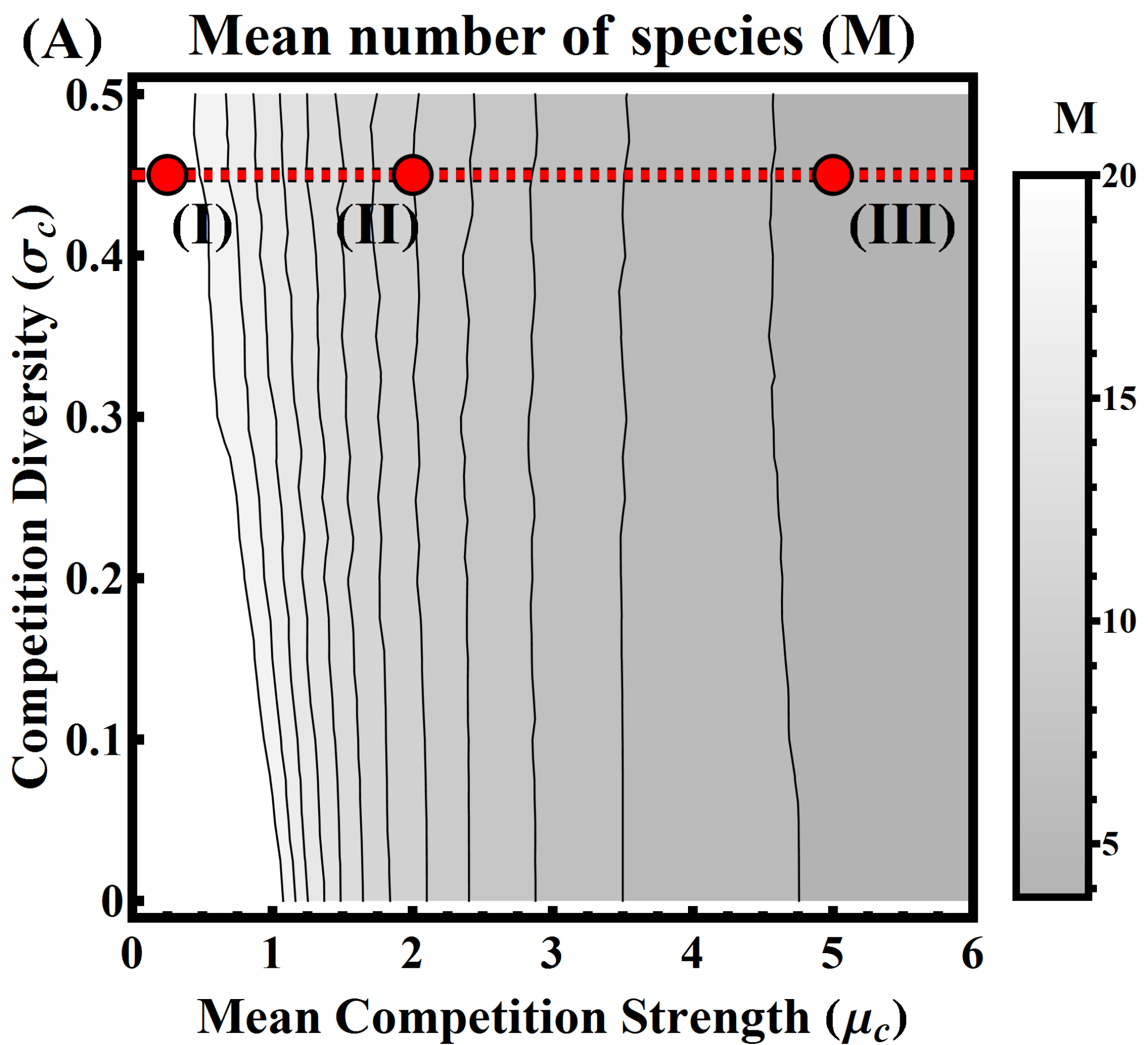} &   \includegraphics[scale=.28]{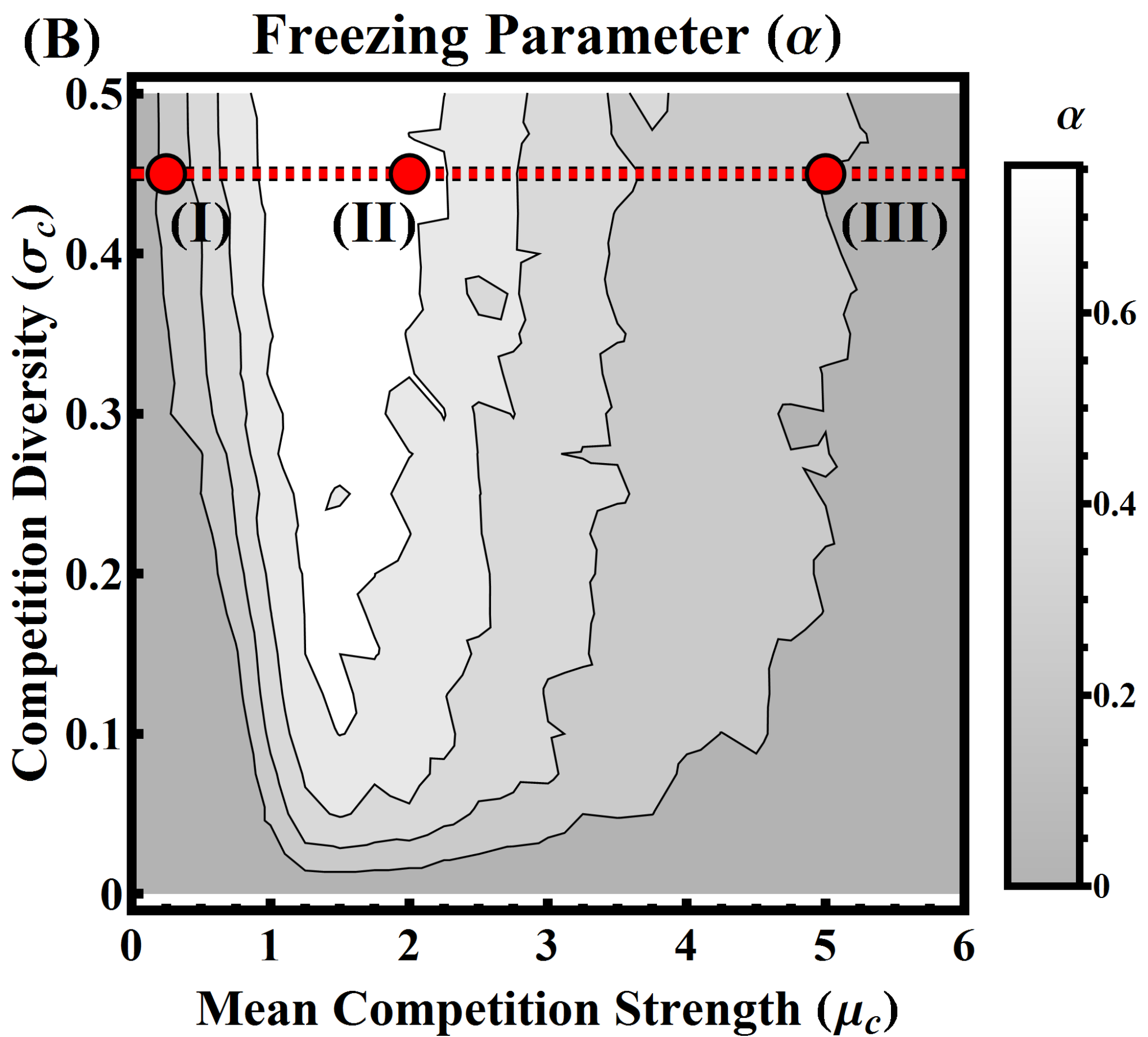}&
 		\includegraphics[scale=.28]{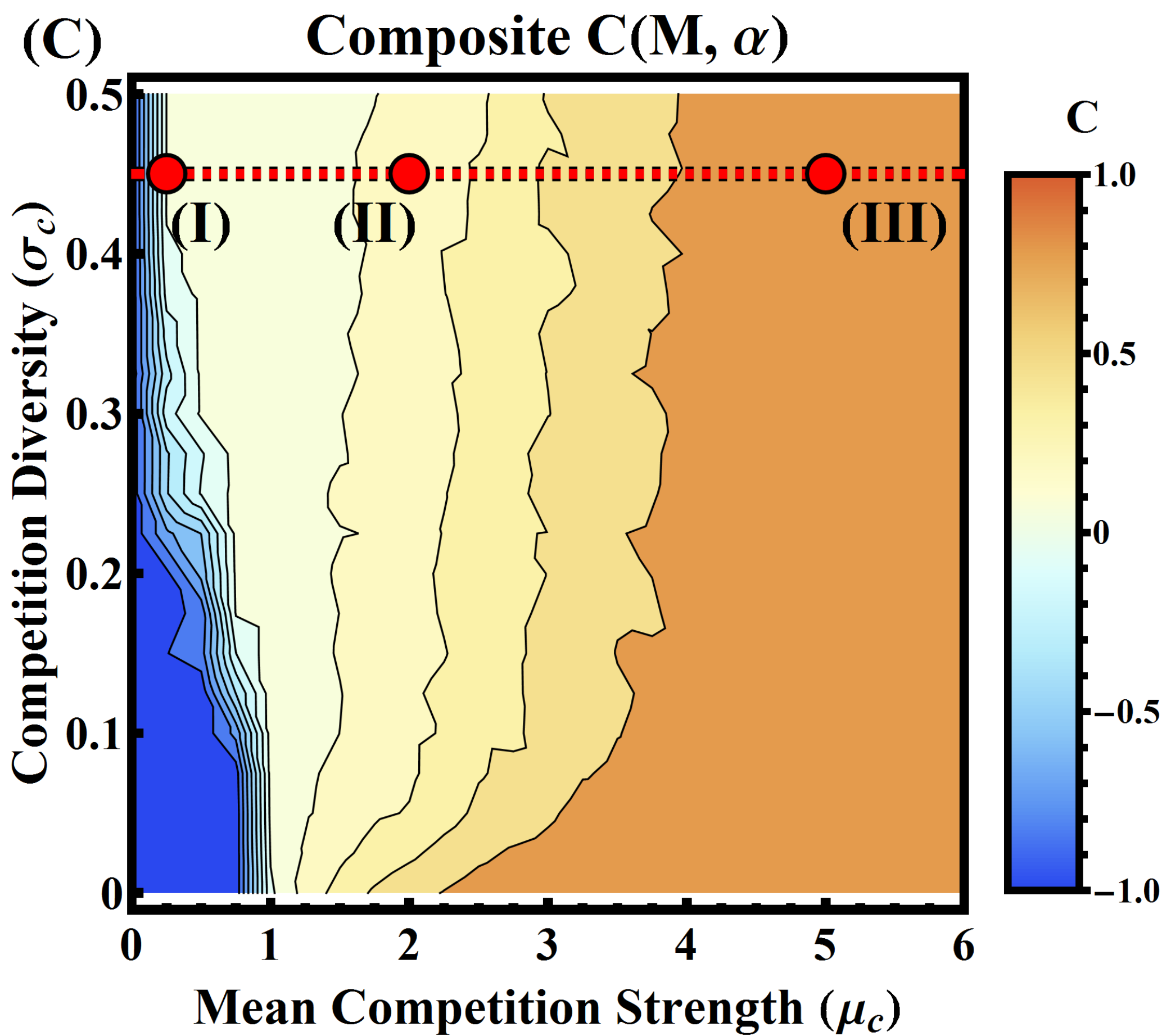} \\ \,&\,&\,\\
 		\includegraphics[scale=.28]{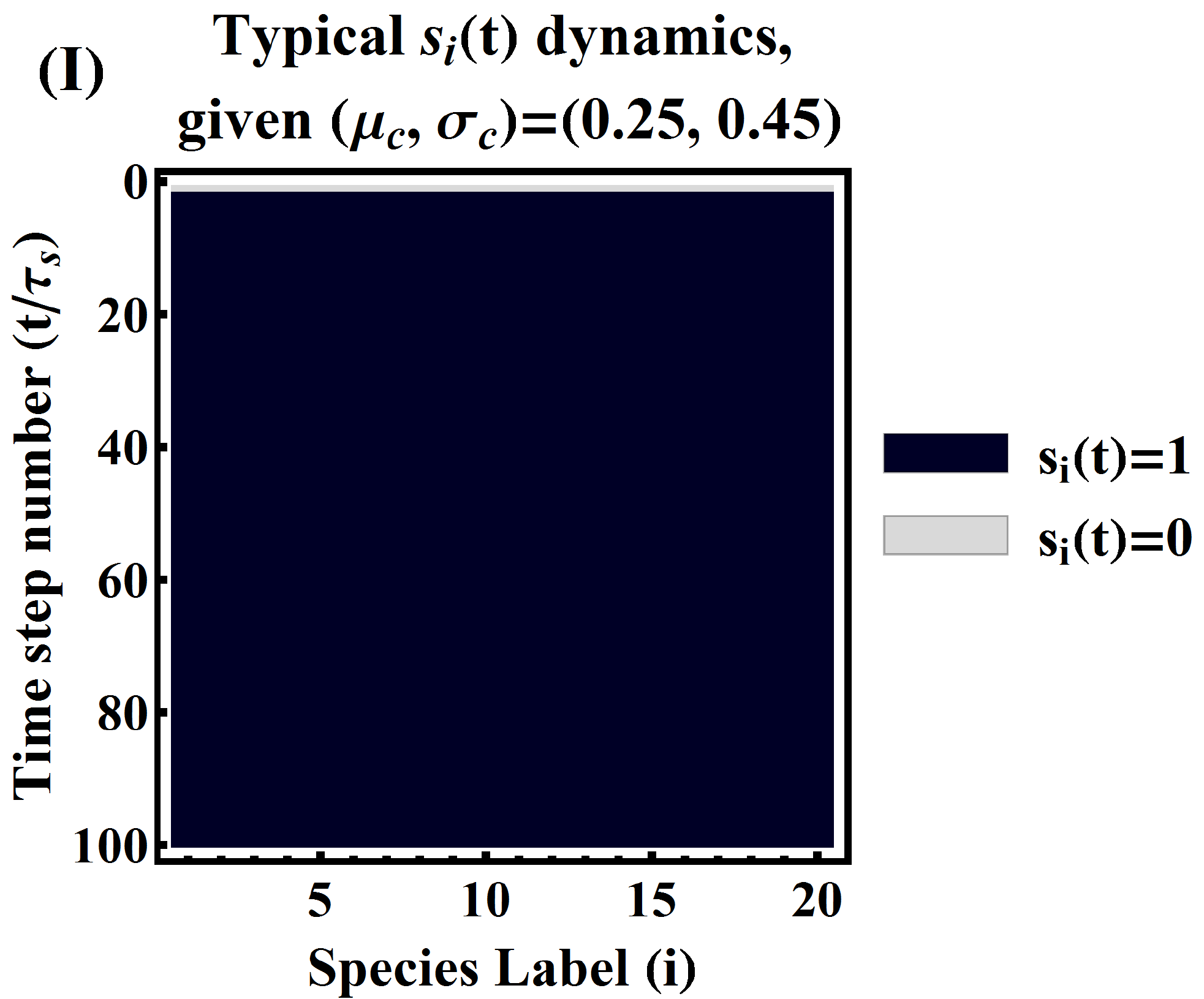} &   \includegraphics[scale=.28]{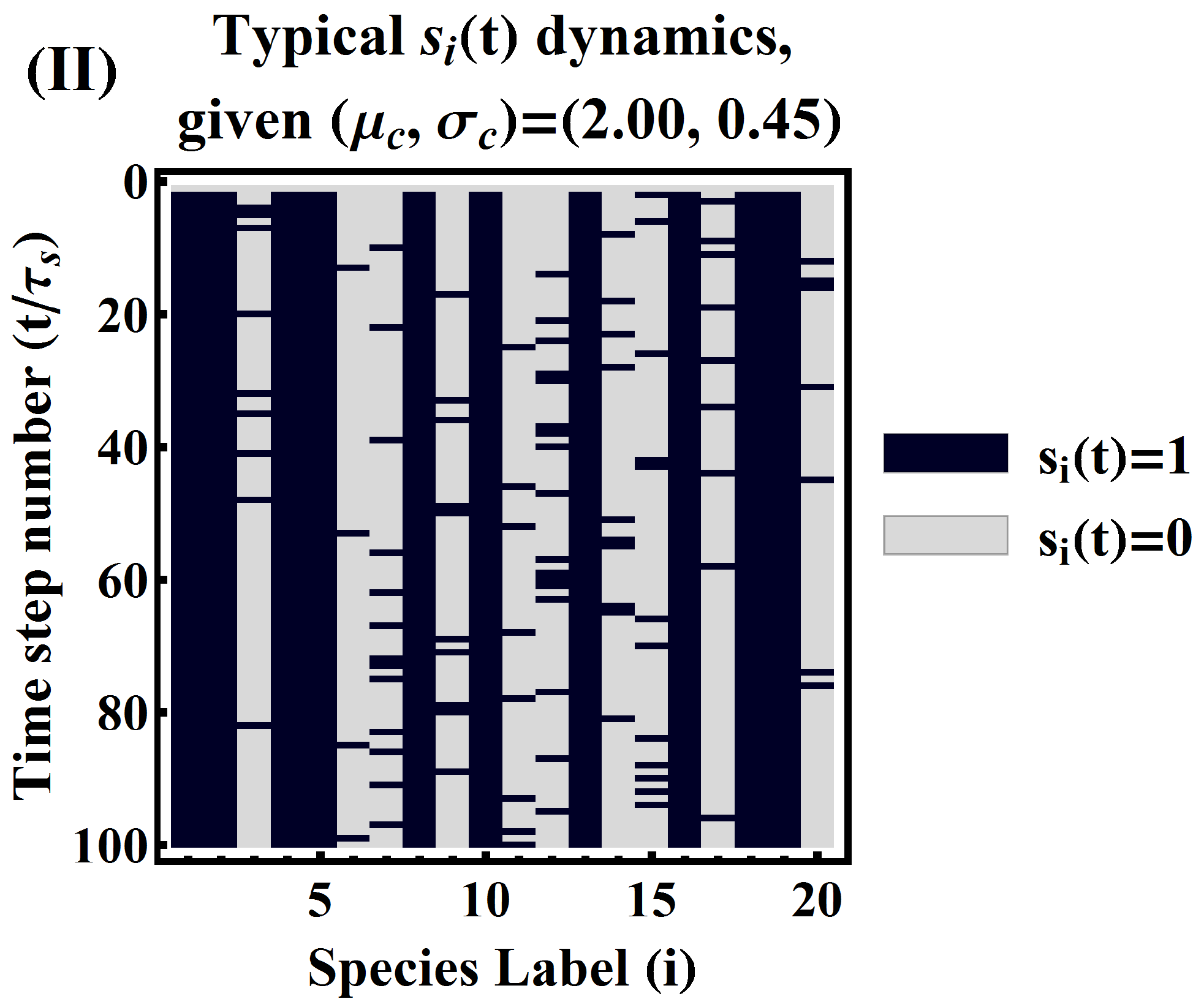} &
 		\includegraphics[scale=.28]{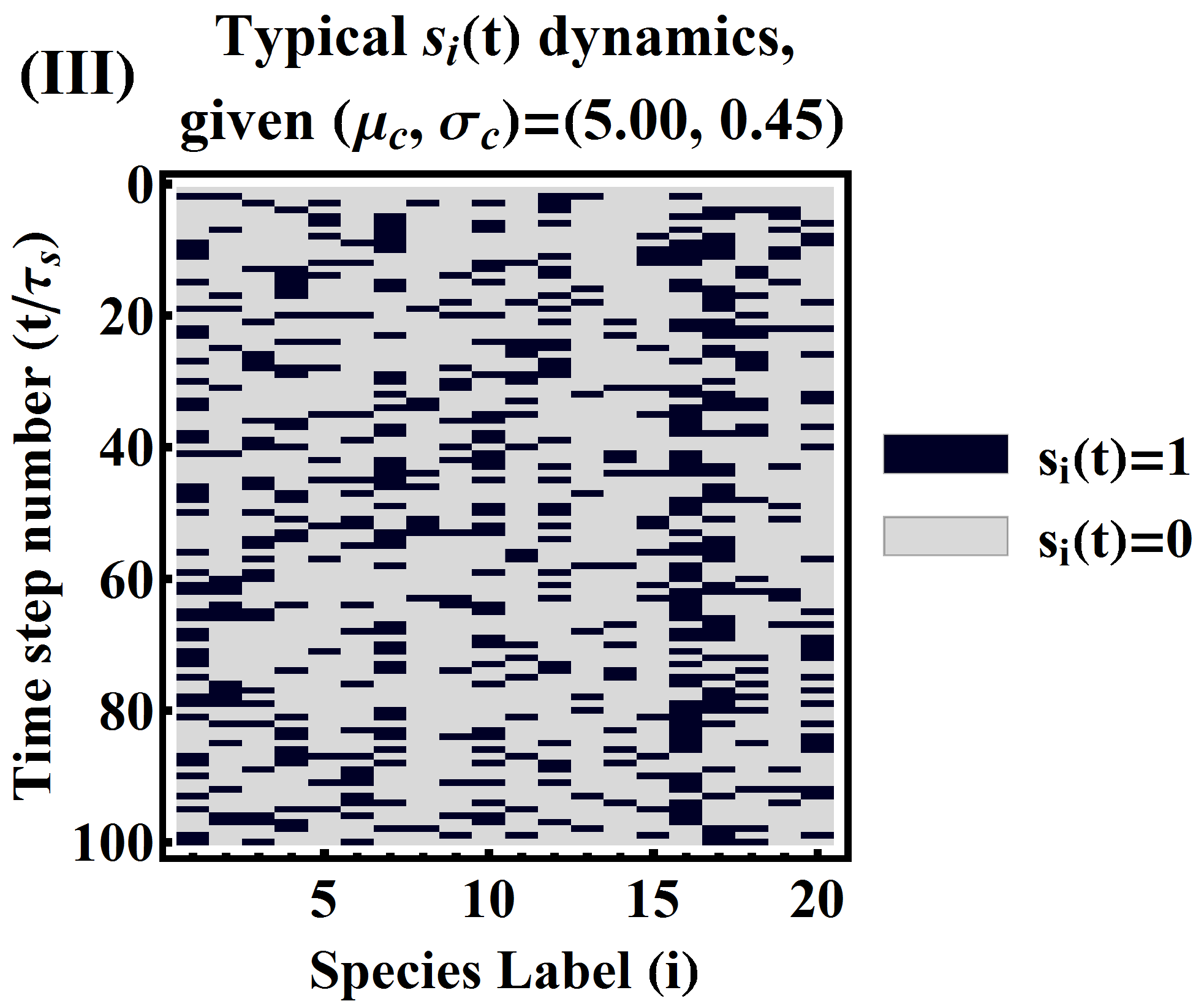}
	\end{tabular}
	 \caption{{\bf (A)} Mean number of species  $M$, {\bf (B)}  the freezing parameter $\alpha$, and {\bf (C)} the composite order parameter $C(M,\alpha)$ (see main text for definitions) computed from numerical dynamics simulations, with $S=20$, $\mu_K=100$, $\lambda=0.01$, $\omega=1$, and $K_i=:\mu_K$ for all $i$. The dynamics exhibit three distinct regimes:  the coexistence regime (CR) with $C(M,\alpha)\simeq -1$, the partial coexistence regime (PCR) with $C(M,\alpha)\simeq 0$, and the noisy regime (NR) $C(M,\alpha)\simeq 1$ . {\bf (I)}, {\bf (II)}, and {\bf (III)}  illustrate the typical dynamics of the CR, PCR, and NR.
  		} 
		\label{phasec}
\end{figure*}

\begin{figure*}[t]
	\centering
	\begin{tabular}{lll}
   		\includegraphics[scale=.28]{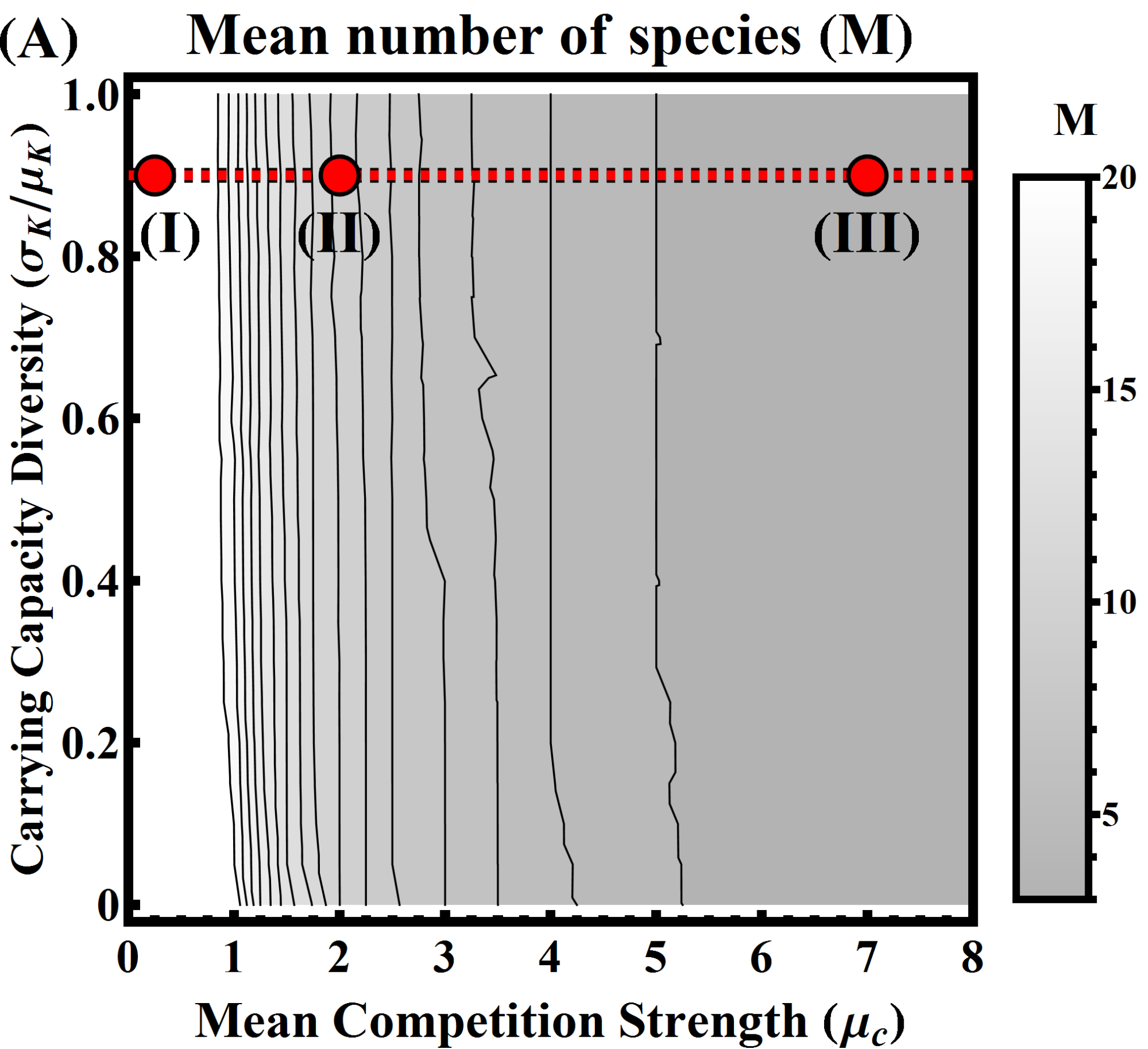} &   \includegraphics[scale=.28]{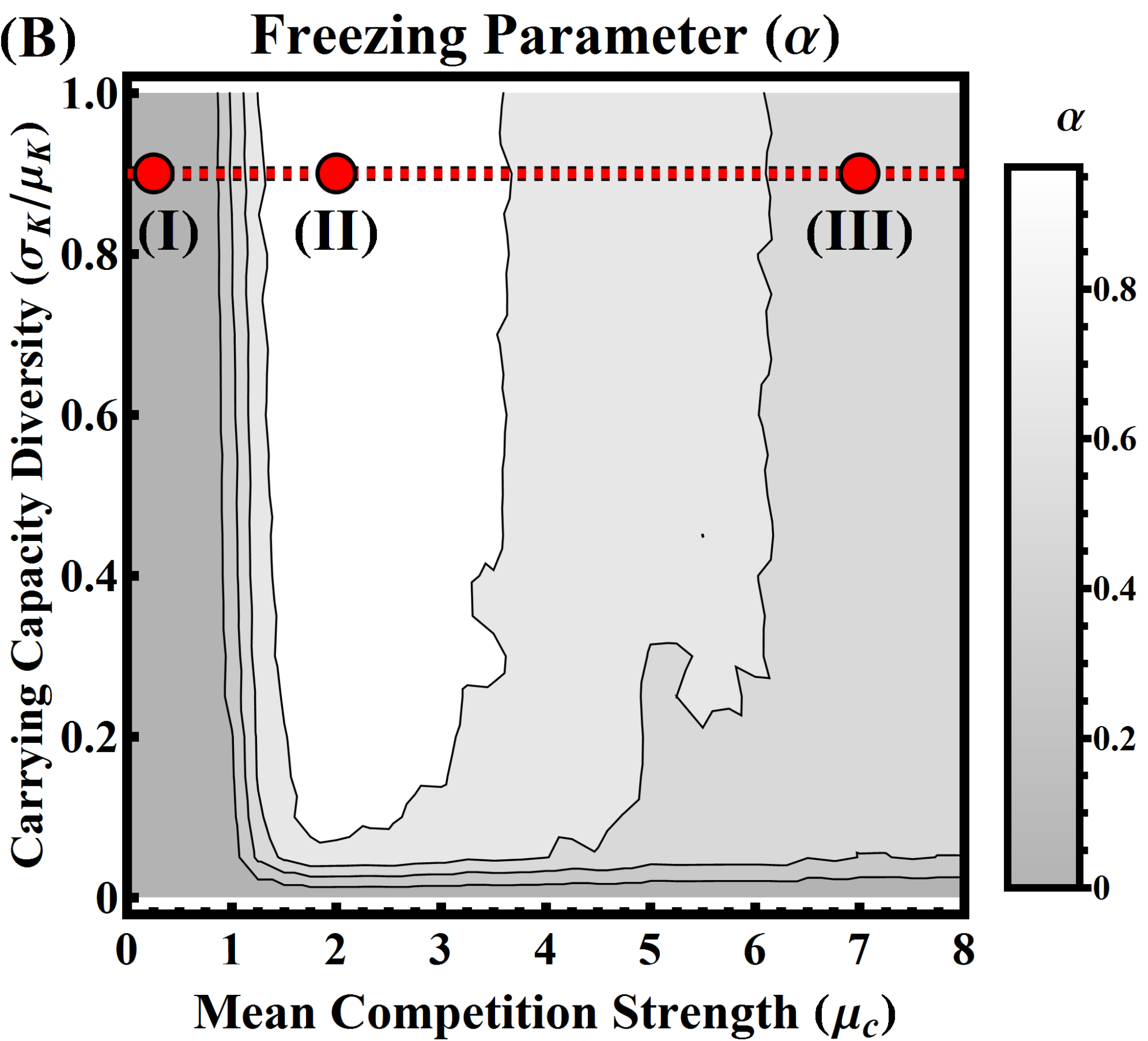}&
 		\includegraphics[scale=.28]{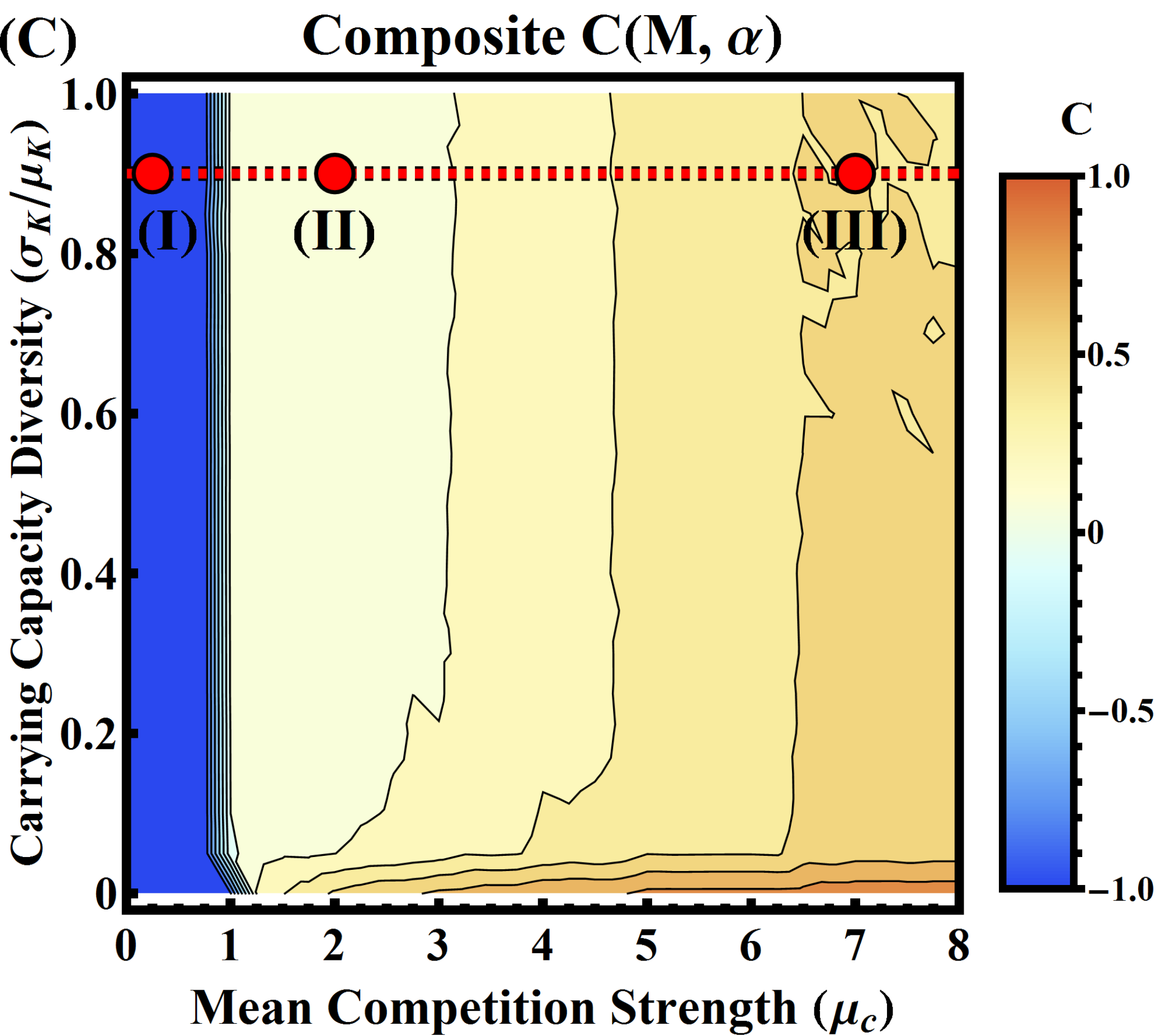} \\ \,&\,&\,\\
 		\,\,\,\,\,\includegraphics[scale=.28]{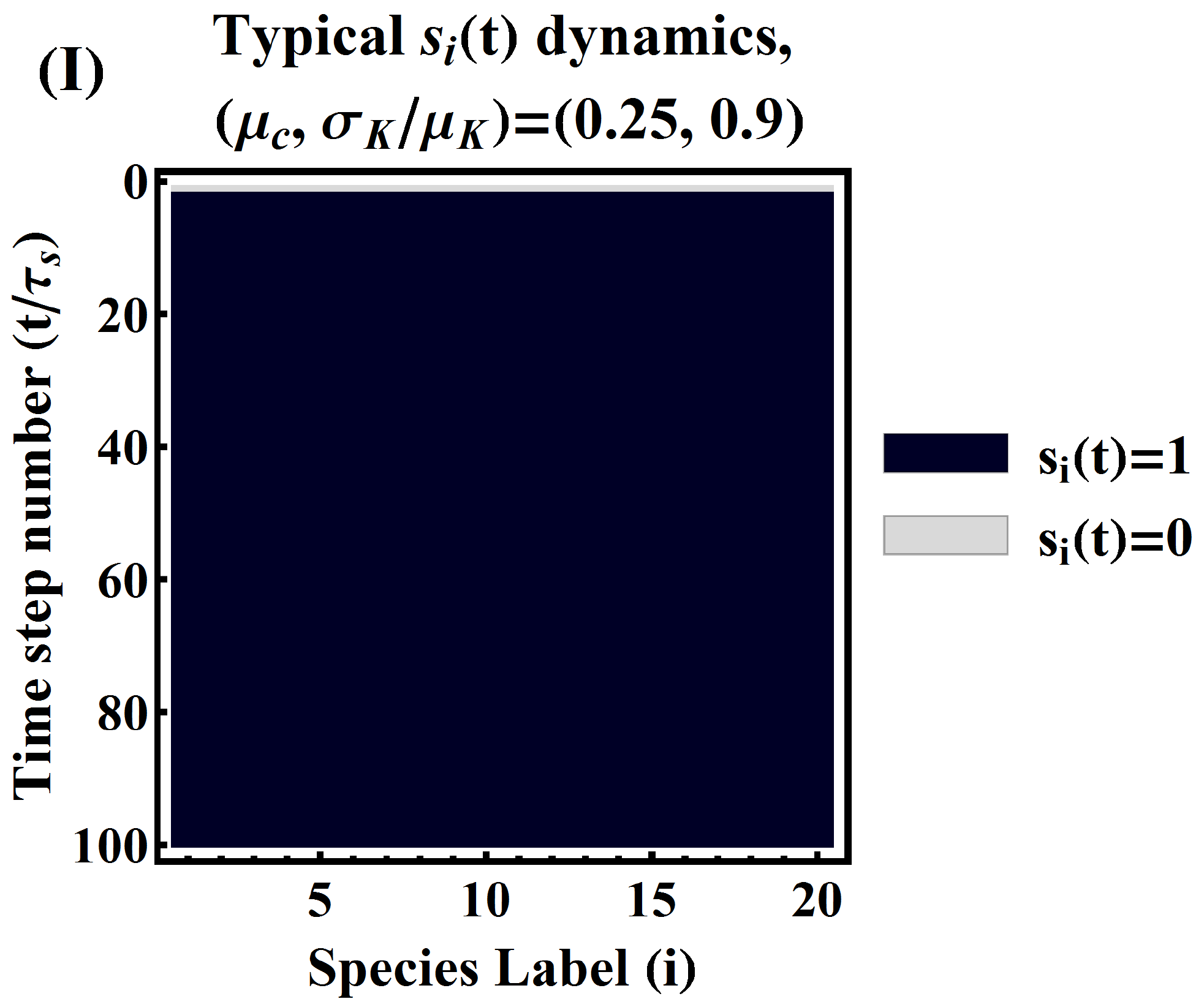} &  
 		\,\,\,\,\,\includegraphics[scale=.28]{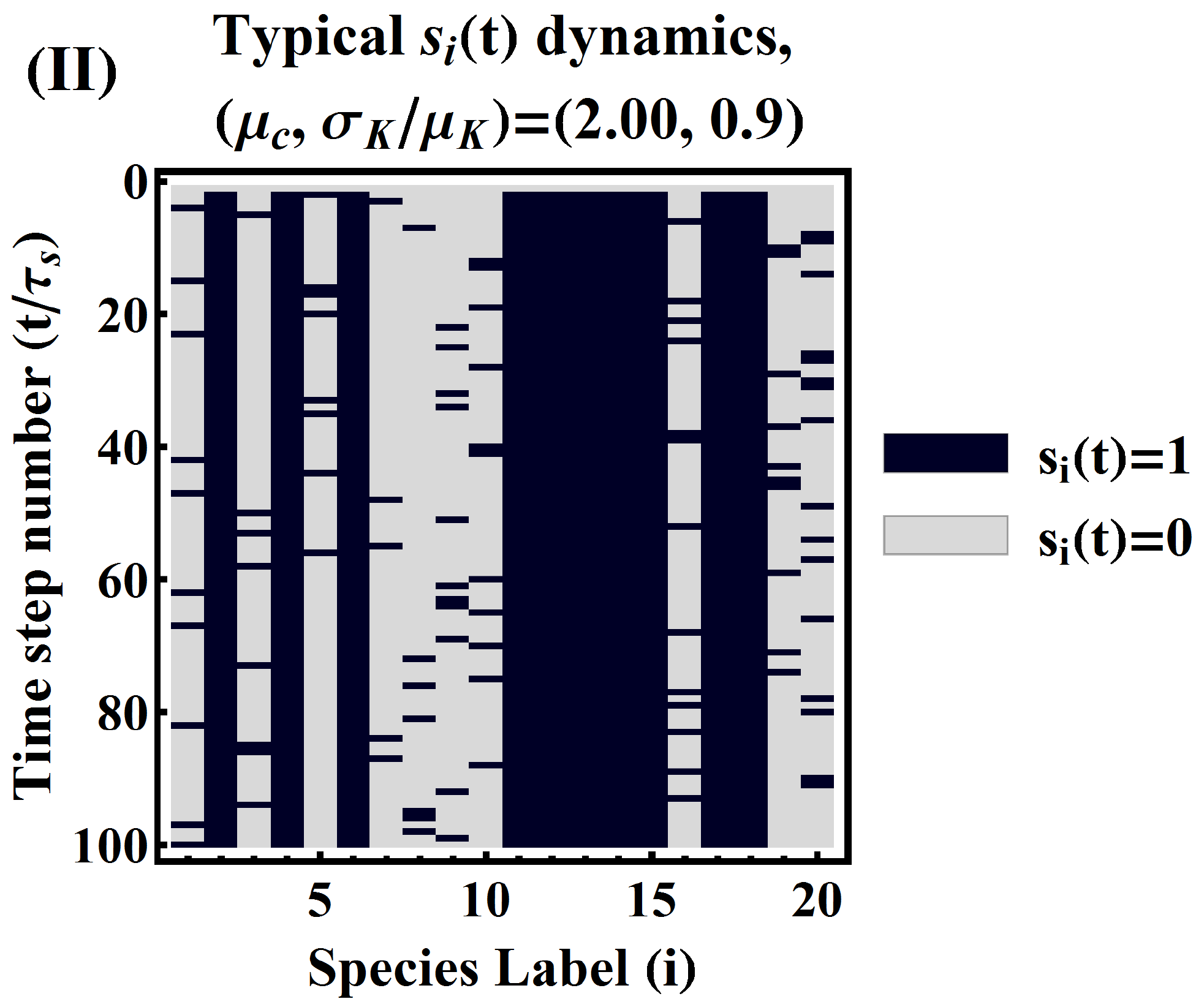} &
 		\,\,\,\,\,\includegraphics[scale=.28]{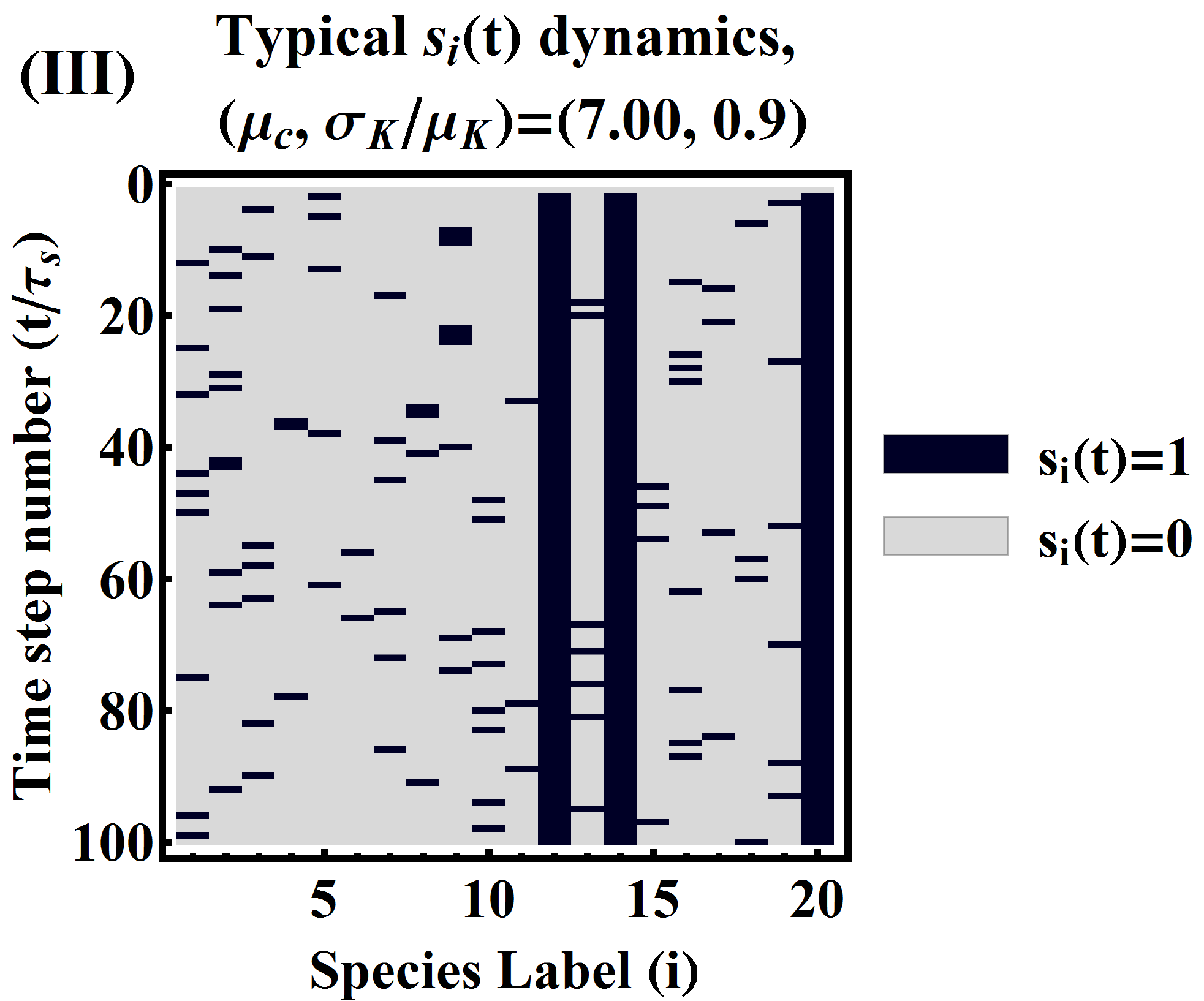}
	\end{tabular}
	 \caption{ {\bf (A)} The mean number of species $M$, {\bf (B)}  the freezing parameter $\alpha$, and {\bf (C)} the composite order parameter  $C(M,\alpha)$ (see main text for definitions) computed from numerical dynamics simulations, with $S=20$, $\mu_K=100$, $\lambda=0.01$, $\omega=0.5$, and $c_{ij}=:\mu_c/S$ fixed for all $i$. For this choice of parameters $\sigma_K/\mu_K=0.9$, the PA Model never reaches the NR until $\mu_c\simeq S$  The dynamics exhibit two regimes:  the coexistence regime (CR) with $C(M,\alpha)\simeq -1$ and the partial coexistence regime (PCR) with $C(M,\alpha)\simeq 0$. {\bf (I)} illustrates the typical dynamics of the CR, whereas {\bf (II)} and {\bf (III)} illustrate the typical dynamics of the PCR. } 
		\label{phaseK}
\end{figure*}

To see if the PA model can reproduce the basic behaviors exhibited by more complicated LVMs \cite{kessler2015generalized,fisher2014transition}, we numerically simulated the PA model dynamics. We found that the dynamics of the PA model can be classified into three broad regimes (see bottom panels of Figures \ref{phasec} and \ref{phaseK}): a coexistence regime (CR) where all species are present,  a partial coexistence regime (PCR) where only a small fraction of species are stably present in the community, and a noisy regime (NR) where all species fluctuate between being present and absent over small timescales. Using order parameters measured in the numerical simulations, we summarized our findings by constructing phase diagrams for these ecological regimes. As seen in \cite{fisher2014transition}, the regimes organize themselves around a special ``critical" point corresponding to Hubbell's neutral theory. We discuss simulation details and results in this section.

\subsection{Simulation details}

In all simulations, we assume that all species have the same immigration rate $\lambda_i=:\lambda$ and ask that $\omega\;|\mathrm{ln}(\lambda)|\ll \mu_K$, where $\mu_K$ denotes the average value of all carrying capacities. Roughly speaking, this assumption assures that the probability for a species to be present or absent is determined primarily by the weakness or strength of that species' extinction rate, $\exp(-K_i^{\mathrm{eff}}(\vec{s})/\omega)$. Thus, in our simulations, the propensity for a species to survive in the local community is determined by its interactions with other species and its environment, as described by the effective carrying capacity, $K_i^{\mathrm{eff}}(\vec{s})$, rather than its immigration ability. 

Numerical simulations of the PA model master equation were performed using Gillespie's algorithm \cite{gillespie1976general}. To compare with the results of \cite{kessler2015generalized}, we started by simulating the PA model for the case where all species have the same carrying capacities $(\mu_K=100,\; \sigma_K=0)$. For each choice of $(\mu_c,\sigma_c)$, 30 random realizations of $c_{ij}$'s were independently drawn from a gamma distribution of mean $\mu_c/S$ and variance $\sigma_c^2/S$. We took $S=20$, $\lambda=0.01$, and $\omega=1$ in these simulations. For each realization, PA Model dynamics were simulated for $T=400,000$ units of PA model time, with $\vec{s}(t)$ sampled in time steps of $\tau_s\simeq 2000$. In addition to heterogeneity in the interaction coefficients, we wanted to investigate the effect of the heterogeneity  in the carrying capacities of species.  Thus, we also performed simulations with random carrying capacities for each choice of $(\mu_c,\sigma_K)$, in which 30 random realizations of the $K_{i}$ vector were independently drawn from a log-normal distribution of mean $\mu_K=100$ and variance $\sigma_K^2$, with $S=20$, $\lambda=0.01$, $\omega=0.5$. Dynamics were simulated for each realization for $T=400,000$ units of PA model time.

\subsection{Order parameters for ecological dynamics}

We constructed phase diagrams for PA model to summarize our findings about its ecological dynamics. Simulations revealed three regimes of qualitatively distinct dynamics: the CR, the PCR, and the NR. These three regimes can be distinguished between by measuring two order-parameter-like quantities from numerical simulation data: the mean number of species and a ``freezing parameter.''  In order to define these quantities, it is necessary to introduce two kinds of averages: time averages, which we denote as  $\langle...\rangle$, and averages over random draws of the ecological parameters $c_{ij}$ or $K_i$, which we denote by $[...]_{av}$.

Define the mean number of species present in the community as 
\begin{equation}
	M:=\sum_{i=1}^{i=S}[\langle s_i\rangle]_{av}.
	\label{defM}
\end{equation}
It is a ``mean" in two senses: it is the number of species averaged both over time and over random draws of species parameters. Intuitively, $M$ tells us whether or not the PA model is exhibiting the coexistence regime. In particular, we expect that $M= S$ in the CR and $M<S$ otherwise. Inspired by the theory of disordered systems, we also define the ``freezing parameter"
\begin{align*}
	\alpha:= \frac{4}{S}\sum_{i=1}^{i=S}\bigg([\langle s_i\rangle^2]_{av}-[\langle s_i\rangle]_{av}^2\bigg).
\end{align*}
$\alpha$ is proportional to the sum of the variances of $\langle s_i\rangle$ over random draws of species traits, and tells us whether or not the PA model is in the PCR. Our intuition for this quantity is as follows. In the PCR, we expect dominant species to emerge that stay in the ecosystem for almost all time. If species $i$ is such a dominant species, then we have $\langle s_i\rangle\simeq 1$. On the other hand, non-dominant species in the PCR will remain absent from the ecosystem for almost all time, hence a non-dominant species $j$ in the PCR will satisfy $\langle s_j\rangle\simeq 0$. Moreover, the subset of species which are dominant depends on the random draw of species parameters $c_{ij}$ and $K_i$. For this reason, if the variance in $c_{ij}$ and $K_i$ is appreciable, then the variance in which species are dominant will also be appreciable; in particular, the variance in $\langle s_i\rangle$ over random draws of species traits will be maximal, and we expect $\alpha\simeq 1$. On the other hand, if the PA model is exhibiting either the CR or NR, then no species is dominating over the others (recall that, in the CR, all species are present and, in the NR, all species are fluctuating between present and absent), regardless of the random draw of $c_{ij}$'s and $K_i$'s. In this case, fluctuations in $c_{ij}$'s and $K_i$'s over random draws will not lead to a non-zero variance in $\langle s_i\rangle$, and the freezing parameter will be close to zero $\alpha \simeq 0$. Thus, $\alpha$  measures whether or not there are dominant species in the ecosystem. For this reason, we expect that $\alpha\simeq 1$ in the PCR and $\alpha\simeq 0$ otherwise. The name ``freezing parameter" comes from the interpretation that, in the PCR, the ecosystem appears ``frozen" or stuck for almost all time in a configuration in which dominant species are present and non-dominant species are absent. In this sense, we can say that $\alpha$ measures whether or not the ecosystem is ``frozen."

To summarize, we expect that $M= S$ and $\alpha\simeq 0$ in the CR, $M<S$ and $\alpha\simeq 1$ in the PCR, and $M<S$ and $\alpha\simeq 0$ in the NR; thus, between these two quantities, we can distinguish between all three regimes from numerical simulation data (see Figures \ref{phasec} and \ref{phaseK}). 

To compare to phase diagrams obtained by analytic calculations and through other models, it is useful to define a composite order parameter $C(M,\alpha)$ whose value distinguishes all three dynamical regimes. We ask that it satisfies $C(M,\alpha)\simeq -1$ in the CR, $C(M,\alpha)\simeq 0$ in the PCR, and $C(M,\alpha)\simeq 1$ in the NR. Moreover, $C(M,\alpha)$ should be continuous and monotonic in both $M$ and $\alpha$. Many functions satisfy these properties. Here we choose
\begin{align*}
	C(M,\alpha):=(1-\alpha)g_{S,\gamma}(M),
\end{align*}
where
\begin{align*}
	g_{S,\gamma}(x):=\left\{
	\begin{array}{c} 
		\displaystyle\frac{\gamma-x}{S-\gamma}, \;\;\mathrm{if}\;\;  x>\gamma,
		\\[1ex]
		\displaystyle\frac{\gamma-x}{\gamma}, \;\;\mathrm{if}\;\; x<\gamma.
	\end{array}
	\right.
\end{align*}
When $S=20$ and $\gamma=19$, $C(M,\alpha)$ will be negative whenever $M>\gamma=19$ and positive if $M<\gamma=19$. With this definition, $C(M,\alpha)$ satisfies the desired properties.

\subsection{The PA model phase diagrams}

In order to understand the effect of competition on the dynamics of the PA model, we constructed a phase diagram as a function of the mean strength of competition ($\mu_c$) and the competition diversity ($\sigma_c$) when all species have identical carrying capacities ($\sigma_K=0$). The results are shown in Figure \ref{phasec}. As expected, the average number of species present in the community ($M$) decreases with increasing competition. Furthermore, for uniform interaction coefficients ($\sigma_c=0$), all species are present in the community until a critical competition strength, $\mu_c=1$, after which species start going extinct.  The middle panel in the figure shows the freezing parameter ($\alpha$), which measures whether a subset of species are consistently in the environment. Notice that this occurs around  $\mu_c \simeq 1$ when the interaction coefficients are heterogeneous. 

Taken together, these numerical observations suggest that the competitive CR is favored when the mean competition strength is low, whereas the NR-type dynamics are favored when competition is very strong. This is consistent with our intuition that a species $i$ should tend to be present if $K_i^{\mathrm{eff}}(\vec{s})>0$ or tend to be absent if $K_i^{\mathrm{eff}}(\vec{s})<0$. To see this, note that $K_i^{\mathrm{eff}}(\vec{s})=\mu_K-\mu_K\sum_{j\not=i}c_{ij}s_j$ in this case. Therefore, $K_i^{\mathrm{eff}}(\vec{s})>0$ for all $i$ precisely when competition is sufficiently low, namely when the CR occurs. On the other hand, we expect to see $K_i^{\mathrm{eff}}(\vec{s})<0$ in the presence of some species when mean competition is high, which is when the NR is observed. In the NR, the dynamics appear to be dominated by stochasticity and drift because species quickly go extinct after immigrating into the community. This can be explained by the fact that negative carrying capacities $K_i^{\mathrm{eff}}(\vec{s})<0$ set a fast extinction rate. At intermediate levels of competition $\mu_c \simeq 1$, and in the presence of heterogeneity in the interaction coefficients, the ecological dynamics are characterized by partial coexistence where only a subset of species remains present in the community. This PCR is consistent with the statement that $K_i^{\mathrm{eff}}(\vec{s})<0$ for {\it some} species, namely the absent ones, and $K_i^{\mathrm{eff}}(\vec{s})>0$ for the present species. Thus, in the PCR regime, some species are more fit for the environment than others, leading to reproducible species abundance patterns. 

We also examined the effect of heterogeneity in carrying capacities on the dynamics of the PA model. To do this, we constructed phase diagrams as function of the carrying capacity diversity (in units of the mean carrying capacity, i.e. $\sigma_K/\mu_K$)  and the mean competition strength for uniform interaction coefficients, assuming that $\sigma_c=0$ (see Figure \ref{phaseK}). The resulting phase diagram once again exhibits three phases with the NR favored when there is strong competition and the CR favored when competition is weak.

A striking aspect of the phase diagrams is that the dynamical regimes organize themselves around a special point in the PA model parameter space where $\mu_c=1$ and $\sigma_{c/K}=0$.  Just as in generalized LVMs \cite{kessler2015generalized}, we can identify this point with Hubbell's neutral theory. To see this, note that all species are equivalent with respect to their ecological traits such as immigration rates, carrying capacities, and competition coefficients at this point. Moreover, the intraspecies competition, described by $K_i$, balances the interspecies competition, given by $\sum_{j\not= i}K_jc_{ij}s_j$. More precisely, if all species are present in the community ($s_i=1$ for all $i$), then the effective carrying capacity of every species is zero, $K_i^{\mathrm{eff}}(\vec{s})=0$. This holds because $c_{ij}=1/S$ for all pairs $i\not=j$, so
\begin{align*}
	K_i^{\mathrm{eff}}(\vec{s})=\mu_K\left(1-\frac{1}{S}\sum_{j\not= i}s_i\right) = \frac{1}{S} \simeq 0
\end{align*}
for $S \gg 1$.  These are precisely the conditions characterizing Hubbell's neutral model \cite{hubbell_unified_2001, volkov_neutral_2003, rosindell_unified_2011,rosindell_case_2012}. Small perturbations around this Hubbell point can lead to qualitatively different species abundance patterns and dynamical behaviors.

\section{Analytic results}

\begin{figure*}[t]
	\begin{center}
	\begin{tabular}{lll}
 		\includegraphics[scale=.28]{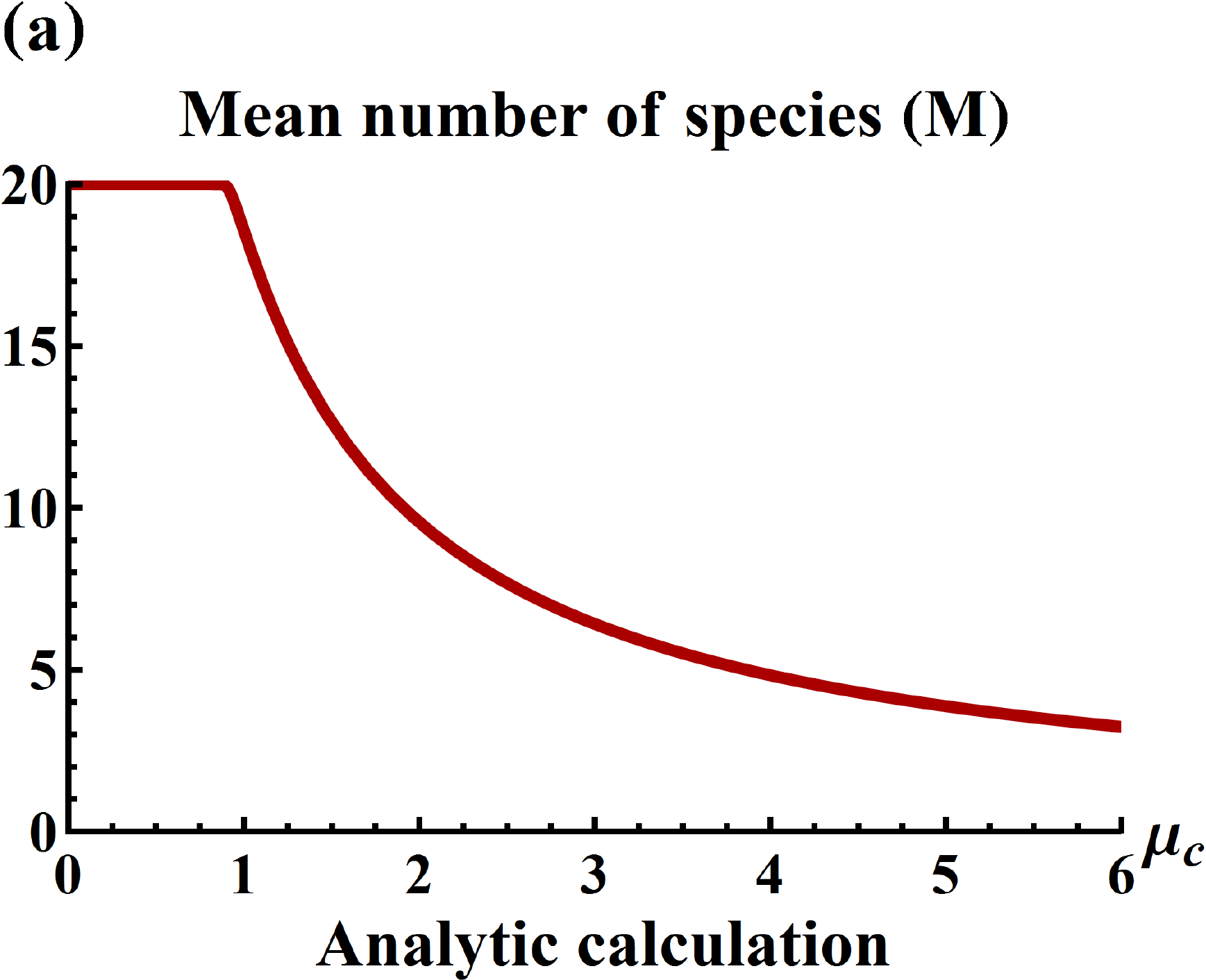} &
 	 	\includegraphics[scale=.28]{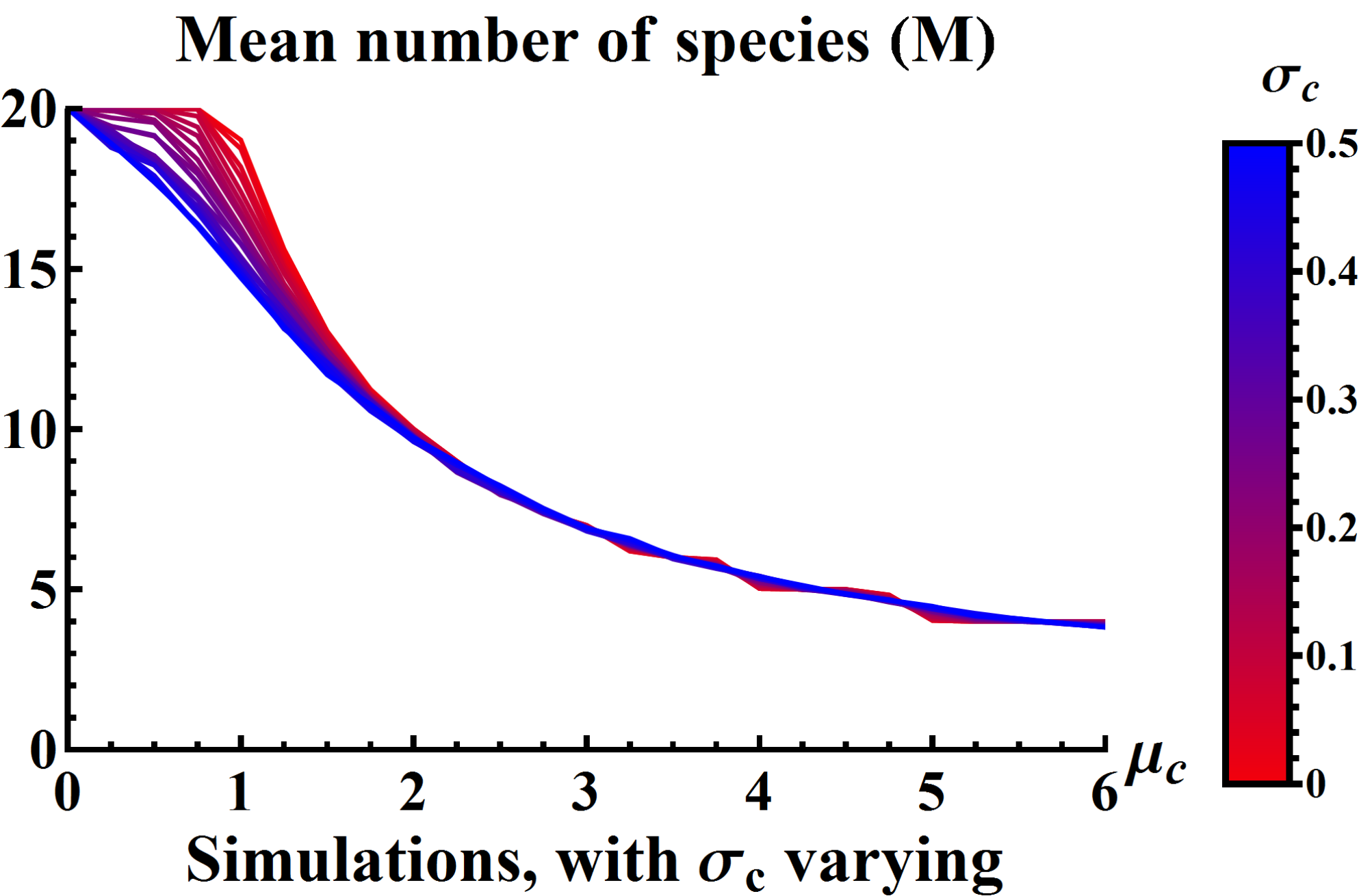} &
 		\includegraphics[scale=.28]{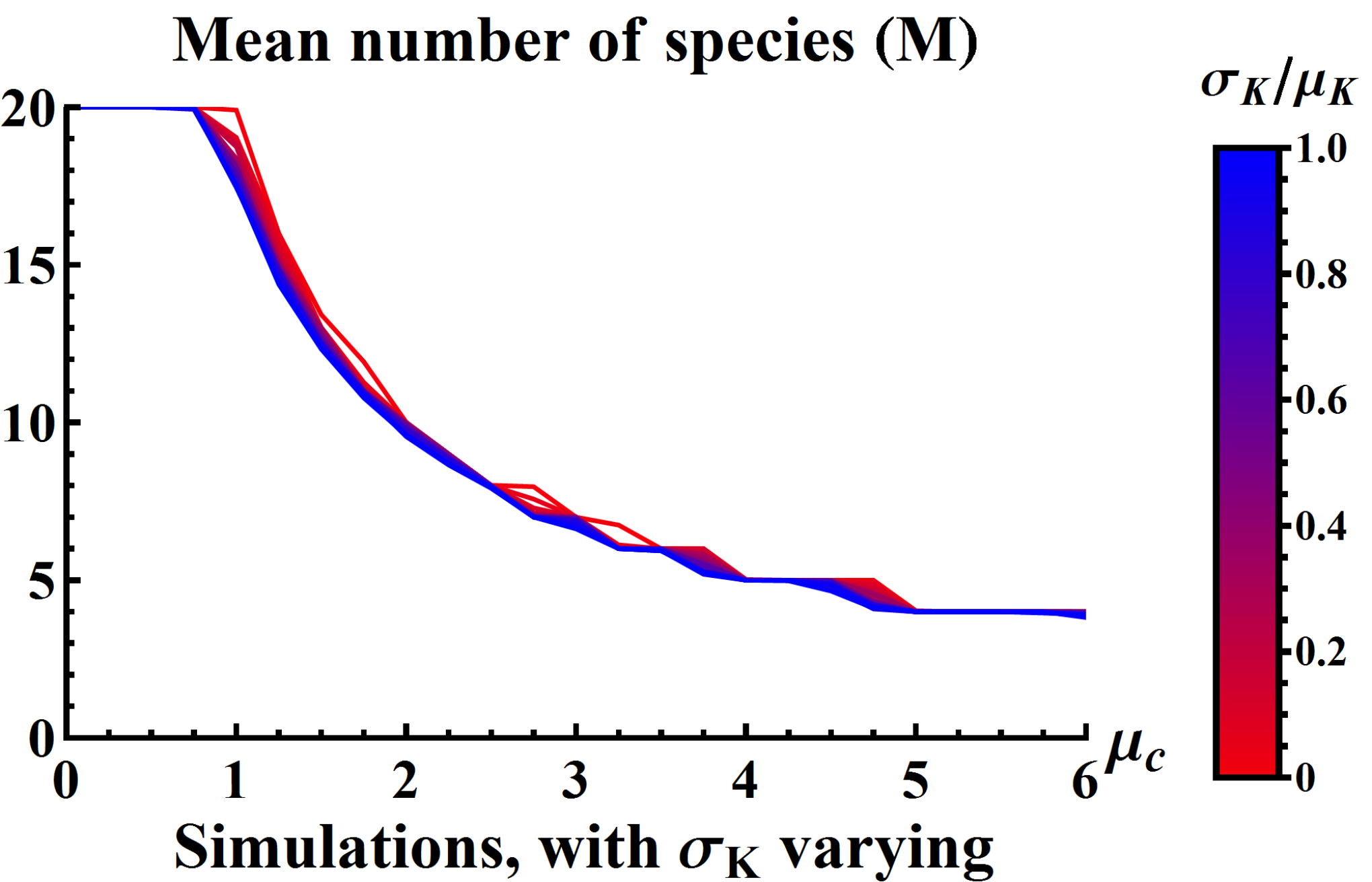} \\
 		&&
	\end{tabular}
	\begin{tabular}{ll}
 		\,\,\,\,\includegraphics[scale=.28]{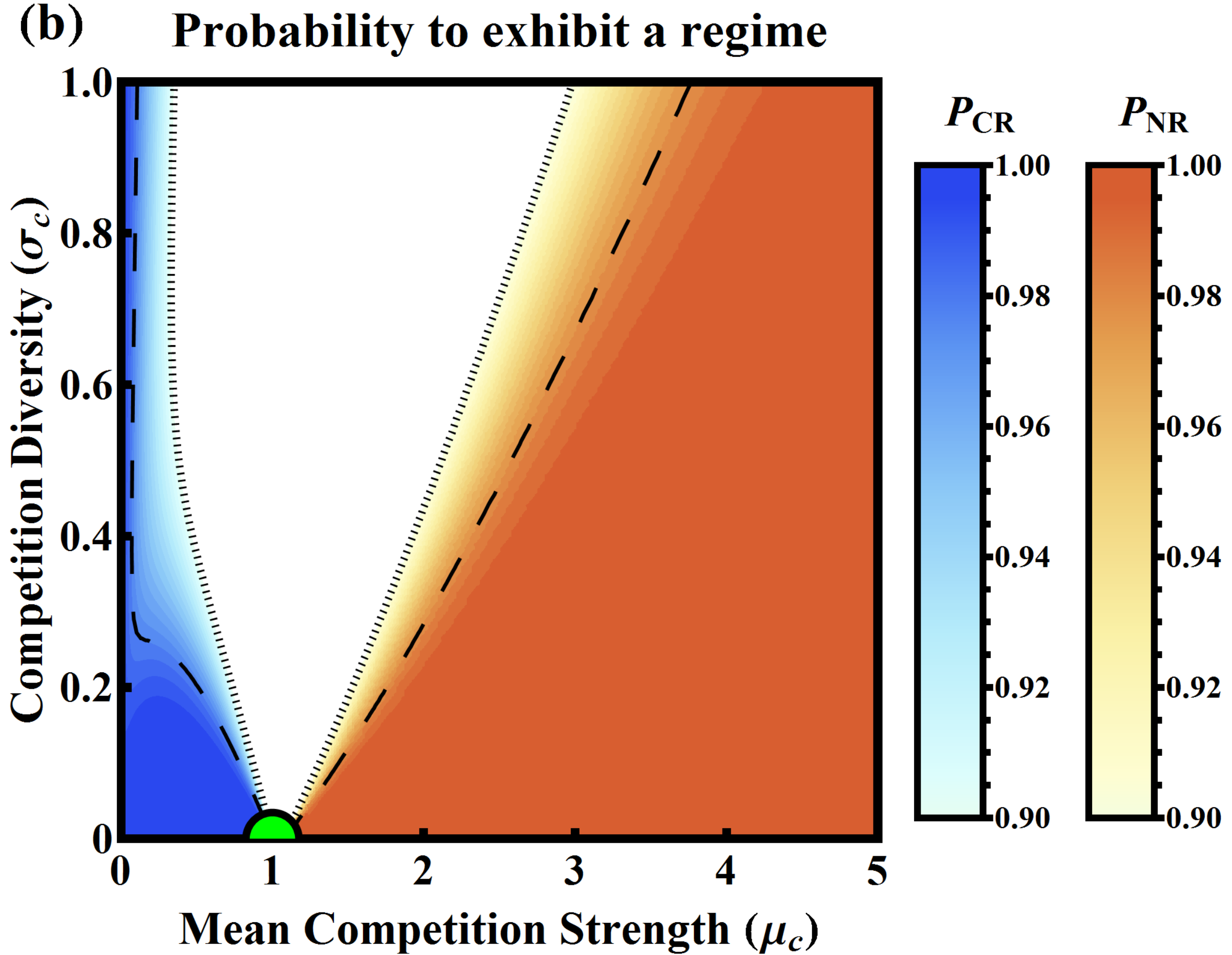}\,\,\,\, \,\,\,\, &
 		 \,\,\,\, \,\,\,\,\includegraphics[scale=.28]{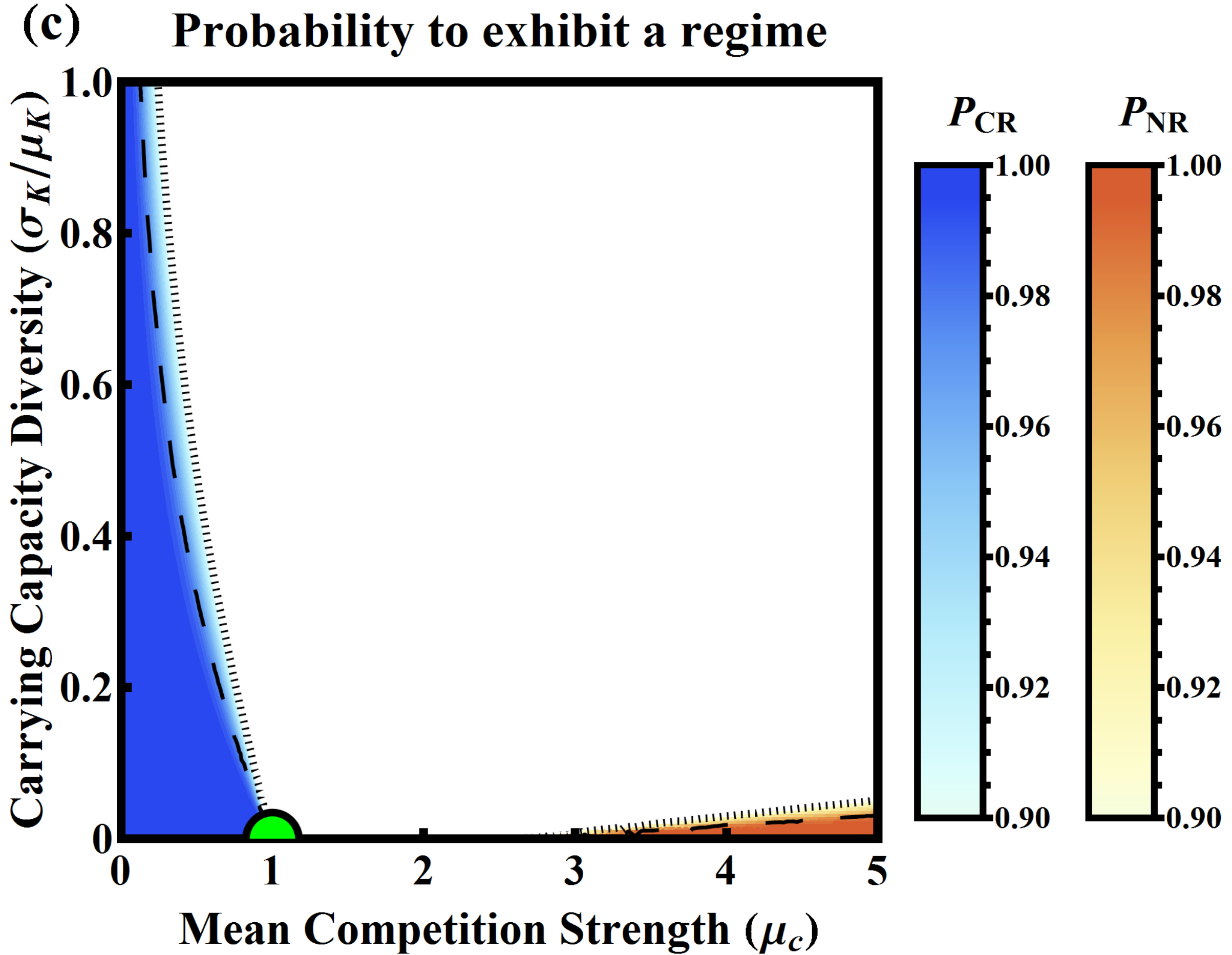} \,\,\,\,
	\end{tabular}
  		\end{center}
  	\caption{
  		{\bf (a)} Mean number of species $M$ computed analytically with $\lambda=.01$, $\omega=1$, and $\mu_K=100$. For comparison, we display $M$ computed numerically, using the data shown in Figures \ref{phasec} and \ref{phaseK}. {\bf (b)} and {\bf (c)}: Probability to exhibit the CR (blue) or the NR (red), as calculated analytically. Long-dashed curves mark where the probability to exhibit a regime is $P=.98$, whereas dotted curves indicate a probability of $P=.90$. In {\bf (b)}, each $c_{ij}$ is independently drawn from a gamma distribution of mean $\mu_c/S$ and variance $\sigma_c^2/S$, while $K_i=\mu_K$ is fixed for all $i$ ($\sigma_K=0$). In {\bf (c)}, each $K_i$ is drawn independently from a log-normal distribution of variance $\sigma_K^2$ and fixed mean $\mu_K$, whereas $c_{ij}=:\mu_c/S$ for all $i\not=j$ ($\sigma_c=0$). In {\bf (b)} and {\bf (c)}, $\lambda=.01$, $\omega=1$, and $\mu_K=100$.
  		}
\label{fig:analytics}
\end{figure*}

In order to better understand our numerical simulations, we performed analytic calculations on the PA model.  Recall that a species $i$ will tend to persist in the local community provided that its extinction rate $\exp(-K_i^{\mathrm{eff}}(\vec{s})/\omega)$ is small (much less than one), and will tend to go extinct if its extinction rate is large (much bigger than one). It follows that a species will probably persist if it tends to have a positive effective carrying capacity, whereas it will go extinct quickly if its effective carrying capacity tends to be negative. This basic observation suggests criteria for classifying the three dynamical regimes analytically in terms of effective carrying capacities. We emphasize that this discussion is relevant only when $\mu_K/\omega$ is large. Since the noise strength, $\omega$, simply provides the units in which to measure $K_i^{\mathrm{eff}}(\vec{s})$, we may take $\omega=1$ here without loss of generality and ask that $\mu_K$ is taken ``large." $\mu_K=100$ is sufficient for our purposes, as in our numerical simulations.

To provide precise criteria for the three regimes in terms of $K_i^{\mathrm{eff}}(\vec{s})$, we introduce a quantity
\begin{align*}
	\kappa_i&=\kappa_i(c,\vec{K})&\\
		&=-\mu_K\mathrm{ln}\bigg\langle \exp\left(-\frac{1}{\mu_K}K_i^{\mathrm{eff}}(\vec{s})\right)\bigg\rangle&\\
		&=\langle K_i^{\mathrm{eff}}(\vec{s})\rangle-\frac{\mu_K}{2}\mathrm{Var}_t\bigg[\frac{1}{\mu_K}K_i^{\mathrm{eff}}(\vec{s})\bigg]+
		\mathrm{h.o.t.c.},&
		\label{defKappa}
\end{align*}
where $\mathrm{Var}_t[...]$ denotes a time-variance and ``h.o.t.c.'' stands for terms proportional to the ``higher order time-cumulants'' of $K_i^{\mathrm{eff}}(\vec{s})/\mu_K$. In statistics language, $\kappa_i$ is proportional to the cumulant generating function of $-K_i^{\mathrm{eff}}(\vec{s})/\mu_K$, where the averages are over time-fluctuations of the species. We sometimes employ the notation $\kappa_i(c,\vec{K})$ to remind ourselves that $\kappa_i$ depends on randomly drawn $c_{ij}$ and $K_i$. 

Our basic intuition about $\kappa_i$ can be summarized as follows. The above equation shows that $\kappa_i$ equals the mean $\langle K_i^{\mathrm{eff}}(\vec{s})\rangle$, minus some cumulant terms that represent the ``typical fluctuations" of the effective carrying capacity. Denote the sum of these cumulant terms by $\delta K_i^{\mathrm{eff}}$, so that $\kappa_i=\langle K_i^{\mathrm{eff}}\rangle-\delta K_i^{\mathrm{eff}}$, and note that $\delta K_i^{\mathrm{eff}}$ is positive-definite due to Jensen's Inequality. By comparing $\langle K_i^{\mathrm{eff}}\rangle$ and $\delta K_i^{\mathrm{eff}}$, we obtain important information about how often $K_i^{\mathrm{eff}}(\vec{s})$ will be positive or negative and, by extension, whether or not we expect species $i$ to survive or go extinct. For example, if $\langle K_i^{\mathrm{eff}}\rangle>\delta K_i^{\mathrm{eff}}$, then we expect that $K_i^{\mathrm{eff}}(\vec{s})$ will tend to fluctuate in the positive real line, hence species $i$ will persist. On the other hand, if $0<\langle K_i^{\mathrm{eff}}\rangle<\delta K_i^{\mathrm{eff}}$, then $K_i^{\mathrm{eff}}(\vec{s})$ fluctuates between being positive and negative, hence species $i$ fluctuates between states of probable persistence and probable extinction. If, instead, $\langle K_i^{\mathrm{eff}}\rangle<0$ and $\langle K_i^{\mathrm{eff}}\rangle<\delta K_i^{\mathrm{eff}}$, then $K_i^{\mathrm{eff}}(\vec{s})$ is almost always negative and species $i$ is almost always absent unless it attempts to immigrate. This intuition suggests the following criteria for the PA model ecological regimes:

\begin{itemize}
\item { (CR)} {\it Coexistence Regime}: The CR occurs when $\kappa_i(c,\vec{K})>0$ for all $i\in \left\{{1,...,S}\right\}$. In this regime, all species tend to coexist stably in the local community because their effective carrying capacities tend to fluctuate in the positive real line, permitting all species to survive. As we will show in the appendix, this criterion reduces approximately to a simpler criterion: the CR occurs when $K_i(\vec{s})>0$ for all $i\in \left\{{1,...,S}\right\}$, when
all species are present. 

\item {(NR)} {\it Noisy Regime}: The NR occurs when $\kappa_i(c,\vec{K})<0$ for all $i\in \left\{{1,...,S}\right\}$. In this regime, the effective carrying capacities of species are either fluctuating between being positive and negative or are fluctuating in the negative real line. In either case, no species should persist in the ecosystem for a very long time because negative effective carrying capacities do not permit them to survive. As a result, all species fluctuate between being present and absent over small timescales, yielding dynamics that appear to be noise-dominated.

In the appendix, we show that this criterion has a natural interpretation in terms of robustness to a typical fluctuation in species abundances. Denote the average number of species in the local community by $M$. Define a new quantity,
\begin{align*}
	\Delta:=\frac{1}{\mu_c\mu_K}\sum_{i=1}^{i=S}[\delta K_i^{\mathrm{eff}}]_{av},
\end{align*}
with $\delta K_i^{\mathrm{eff}}=\langle K_i^{\mathrm{eff}}\rangle-\kappa_i$  a ``typical fluctuation'' of the effective carrying capacity for species $i$. In the appendix, we show that the a system will be in NR if $K_i(\vec{s})<0$ for all $i\in \left\{{1,...,S}\right\}$, whenever $M+\Delta$ species are present, where $M$ is the mean species abundance defined in (\ref{defM}). In particular, when the number of species fluctuates a threshold fluctuation $\Delta$ above the mean, no species should persist for very long because their carrying capacities become negative.

\item{(PCR)} {\it Partial Coexistence Regime}: This occurs when $\kappa_i(c,\vec{K})<0$ for a {\it fraction} of species $i\in \left\{{1,...,S}\right\}$ and $\kappa_i(c,\vec{K})>0$ for other species $i$. In this regime, we expect dominant species to emerge in the community, namely those with positive $\kappa_i(c,\vec{K})$, and remain in the ecosystem for almost all time. Meanwhile, those species with negative $\kappa_i(c,\vec{K})$ will attempt and fail to invade, going extinct quickly and fluctuating between presence and absence. \\
\end{itemize}

Given these criteria, one can use a mean-field-theoretic approach to analytically calculate $M$ and a phase diagram for the PA model (see appendix for details). The results are shown in Fig. \ref{fig:analytics}. To determine the phase boundary of the CR, we calculated the probability, $P_{\mathrm{CR}}$, that $\kappa_i(c,\vec{K})>0$ when $M=S$ species are present in the community, given random draws of $c_{ij}$ or $K_i$ (see appendix). This is plotted in the blue region of the left hand side of the analytic phase diagrams in Fig. \ref{fig:analytics}. To determine the boundary of the NR, we calculated the probability $P_{\mathrm{NR}}$ that $\kappa_i(c,\vec{K})<0$. The results of these calculations are shown in the red region of the right hand side of the analytic phase diagrams in Fig. \ref{fig:analytics}. The PCR occurs in the white region where neither the CR nor the NR is probable. 

In the NR, the statistics of species abundances appear to be ``neutral'' (``statistically neutral'' in the language of \cite{fisher2014transition}). Thus, we assumed that species are statistically independent and neglected contributions from the heterogeneity in $c_{ij}$ and $K_i$. The latter assumption implies that an equilibrium probability distribution will be reached as $t\rightarrow\infty$ and that all species have the same mean value $m=\langle s_i\rangle$. The former assumption allows us to write a species probability distribution $Q(\vec{s})=\prod_{i=1}^{i=S}Q_i(s_i)$ in equilibrium that factorizes into individual probability distributions $Q_i(s_i)$. We approximate these marginal distributions, $Q_i(s_i)$, as Gaussian distributions with mean $m$ and variance $\sigma_m^2=m(1-m)$ by neglecting cumulants of $s_i$ of higher order than the variance. This yields the same answer one would obtain by neglecting the moments of $c_{ij}$ and $\frac{1}{S}K_j$ of third and higher order. Similar approximations were also employed to compute the CR phase boundary, but no Gaussian approximation was needed. See the appendix for details.

The agreement between analytic and numerical results is remarkable. The mean-field calculations of $M$ agree with numerical simulations, even for moderately large values of $\sigma_c$ and $\sigma_K$. Despite our approximations, there is a surprising agreement between our analytic phase diagrams and the numerical phase diagrams in Figures \ref{phasec} and \ref{phaseK}. Strikingly, in the case of heterogeneous $K_i$, the NR is very small in both analytic or numerical calculations, where $S=20$. Overall, our results suggest that we can capture the essential ecology of the PA model by thinking about the means and typical fluctuations of the effective carrying capacities of individual species. 

\section{Discussion}

We analyzed the binary, presence-absence (PA) model for community assembly first introduced by \cite{fisher2014transition}. The PA model describes an immigration-extinction process in which species are treated as stochastic binary variables that can either be present or absent in a community. Species immigrate to the community from a regional species pool. Once in the local community, a species competes for resources until it becomes locally extinct due to competition and stochasticity. Here, we investigated the effects of heterogeneous competition coefficients and carrying capacities on the ecological dynamics in large, ``typical'' communities. We found that the PA model exhibits three distinct regimes: a coexistence regime (CR) where all species are present in the community, a noisy regime (NR) where all species quickly go extinct after immigrating to the community leading to neutral-like dynamics, and a partial coexistence regime (PCR) where a broad distribution of effective carrying capacities leads to a few dominant species that remain present in the community most of the time. 

These three regimes all converge at a special point (called the Hubbell point) in parameter space corresponding to the neutral theory of biodiversity, where all species are identical and their dynamics are uncorrelated. The Hubbell point plays an analogous role to a quantum critical point in the phase diagrams of systems that exhibit phase transitions \cite{sachdev2007quantum}. In the absence of heterogeneity in the interaction coefficients or carrying capacities (i.e., $\sigma_c=0$ and $\sigma_K=0$), the Hubbell point separates a selection-dominated regime, where all species are present and the dynamics look fairly deterministic, from a drift-dominated regime, where selection is not important for the dynamics. For non-zero $\sigma_c$ and $\sigma_K$, the effect of the Hubbell point manifests itself in the existence of the partial coexistence regime wherein a subset of the species are always present in the community due to selection, while the dynamics of the remaining species are dominated by noise. This suggests that the neutral theory of biodiversity plays a special role in understanding ecological dynamics, perhaps as much as critical points play an important role in the theory of phase transitions. One interesting question worth investigating is whether ideas such as universality and critical exponents can also be exported to this ecological setting.

Despite its simplicity, the PA model is able to reproduce the qualitative behaviors of more complicated generalized Lotka-Volterra models (LVMs). For example, our phase diagram for the PA model in Figure \ref{phasec} is almost identical to the phase diagram of the LVM obtained using numerical simulations in \cite{ kessler2015generalized}. Nevertheless, while the PA model exhibits three phases, the LVMs were found to exhibit four. Both the PA model and LVMs exhibit coexistence and partial coexistence regimes, respectively at low and intermediate levels of competition. However, instead of a noisy regime at high competition, \cite{ kessler2015generalized} identified a ``disordered'' phase and a ``glass-like" phase. The disordered phase appears to be analogous to our noisy regime. That is, there is no longer a fixed set of resident species which are always present; instead, there is a constant turnover in the community composition. The glass-like phase, which appears at levels of competition greater than that of the disordered phase, is characterized by occasional noise-induced transitions between a few equilibria, such that for each equilibrium only a few dominant species are present. We did not identify this behavior in the PA model. This discrepancy is likely due to the simplified dynamics in the PA model that ignores species abundance distributions. Thus, the PA model provides a compromise between complexity and interpretability given that it is amenable to analytic techniques.

The idea of an effective carrying capacity plays a central role in the PA model. The importance of this quantity was already noted in the early works of Macarthur and Levins \cite{macarthur_limiting_1967}. The effective carrying capacity essentially sets the extinction time in the local community, and measures how susceptible a species is to stochastic events that can cause it to die out. Our analytic calculations demonstrate that a mean-field like picture based on effective carrying capacities is sufficient to reproduce the numerical phase diagram. This suggests that, in large communities with many species, the effect of different ecological processes can be understood by asking how they change the effective carrying capacity for a typical species configuration. This is similar in spirit to recent work in the theoretical ecology literature \cite{chesson2000mechanisms}. These simplifications suggest the behaviors of large ecosystems with many species may differ significantly from the behavior of small systems with a few species.

In this work, we limited ourselves to considering purely competitive interactions in a spatially well-mixed population with low immigration rates from a regional species pool. It will be interesting to generalize these results to the case where the interaction coefficients can be mutualistic, or even hierarchical \cite{bascompte2003nested}. Another important avenue for future research is to ask how the introduction of spatial structure differs from the mean-field picture. In particular, it will be interesting to understand if the phase diagram of the PA model is still organized around Hubbell's neutral theory and if the PA model can reproduce the species-area relationships seen in real ecosystems \cite{rosindell2007species}. 

\section{Acknowledgements:} This work was partially supported by a Simons Investigator in the Mathematical Modeling of Living Systems and a Sloan Research Fellowship to PM. BD also acknowledges the Boston University Undergraduate Research Opportunities Program for partial funding.

\appendix

\section{Mean field approximation and calculating the mean species abundance}

We can analyze the  PA model in the coexistence regime (CR) and the noisy regime (NR) using  mean field theory (MFT). In MFT, the true distribution of species is approximated by an equilibrium variational distribution, $Q(\vec{s})$, that factorizes over species:
\begin{align*}
	Q(\vec{s})=\prod_{i=1}^{i=S}Q_i(s_i).
\end{align*}
For the PA model where $s_1=0$ or $1$, the mean-field variational distribution takes the form
\be
	Q_i(s_i)=m\delta_{s_i,1}+(1-m)\delta_{s_i,0} \label{CRvar},
\ee
where $ \delta_{s_i,1}$ is the Kronecker delta function and  $m$ is a variational parameters which measures the probability of a species being present:  $\langle s_i\rangle=:m$. Notice that we use the same parameter of $m$ for all $i$. 

In the coexistence regime, we know that all species are present so that $m \simeq 1$. For this reason, in the CR the mean field variational ansatz is well approximated by
\be
Q_i^{\mathrm{CR}}(s_i)=m\delta_{s_i,1}
\ee
In the noisy regime, we approximate  $Q_i(s_i)$ by a Gaussian distributions with mean  $m$ and variance $\sigma_m^2=m(1-m)$:
\be
	Q_i^{\mathrm{NR}}(s_i)= \frac{1}{\sqrt{2\pi\sigma_m^2}}\exp\left(-\frac{1}{2\sigma_m^2}[s_i-m]^2\right). \label{NRvar}
\ee
$Q_i^\mathrm{NR}$ can be thought of as an approximation to the full variational distribution $Q_i$ where we have ignored higher order cumulants beyond the variance. Since $m \ll 1$ in the NR, this is expected to be a good approximation.  

These mean field variational ansatz are consistent with numerical simulations of the CR and the NR that show that the heterogeneity of $c_{ij}$ and $K_i$ do not significantly modify the dynamics in these regimes and species appear to be statistically independent because the dynamics of different species are uncorrelated in time.

We would like to compute $m$ for the variational distribution (\ref{CRvar}), which requires some knowledge of the true equilibrium distribution in the absence of heterogeneity ($\sigma_c=\sigma_K=0$). In this case, the dynamics approach a unique equilibrium distribution as $t\rightarrow\infty$ \cite{fisher2014transition}:
\begin{align*}
	P_{PA}(\vec{s})=\frac{1}{Z}\exp\left(-\frac{\mu_K}{\omega}U(\vec{s})\right),\\
	U(\vec{s}):=-[1+\Lambda]\sum_{i=1}^{i=S}s_i+\frac{\mu_c}{2S}\sum_{i\not=j}s_is_j,
\end{align*}
where $\Lambda:=\omega\,\mathrm{ln}(\lambda)/\mu_K$ and $Z$ is a normalization constant. The function $U(\vec{s})$ is sometimes called the ``internal energy,'' or just the ``energy,'' associated with the species configuration $\vec{s}$. Due to ergodicity,  we reinterpret time averages $\langle ...\rangle$ as averages over the distribution $P_{PA}(\vec{s})$. In particular, given a function $A=A(\vec{s})$ of the random variable $\vec{s}$, we have
\begin{align*}
	\langle A\rangle=\mathrm{Tr}_{\vec{s}}\,A(\vec{s})P_{PA}(\vec{s})
	=\frac{1}{Z}\mathrm{Tr}_{\vec{s}}\,A(\vec{s})\exp\left(-\frac{\mu_K}{\omega}U(\vec{s})\right),
\end{align*}
where $\mathrm{Tr}_{\vec{s}}$ denotes the operation of summing over all possible $2^S$ configurations of $\vec{s}$. Since $U(\vec{s})$ is invariant under the exchange $i\leftrightarrow j$ of species indices, it follows that $\langle s_i\rangle=\langle s_j\rangle$ for all $i\not=j$.

With these observations, we can use the functional form (\ref{CRvar}) for $Q(\vec{s})$ to calculate the mean species abundance, $M$. In order to determine the variational $m$, we minimize the variational free energy:
\begin{align*}
	F[Q]=\langle U(\vec{s})\rangle_{Q}+\frac{\omega}{\mu_K}\langle\mathrm{ln}\,Q(\vec{s})\rangle_{Q},
\end{align*}  
where $\langle ...\rangle_{Q}$ denotes an average with respect to $Q$. Since $Q$ is completely specified by $m$, we can express $F[Q]$ as a function $F(m)$ of $m$. One obtains 
\begin{align*}
	\frac{F(m)}{S}=-[1+\Lambda]m&+\frac{\mu_c m^2}{2}&\\
	&+\frac{\omega}{\mu_K}[m\,\mathrm{ln}(m)+(1-m)\,\mathrm{ln}(1-m)].&
\end{align*}
A necessary condition for minimization is that $\frac{d}{dm}{F(m)}=0$, which yields 
\begin{equation} \label{mfe}
	m=\frac{1}{2}+\frac{1}{2}\tanh\left(\frac{\mu_K}{2\omega}\bigg[1+\Lambda-\mu_c m\bigg]\right).
\end{equation}
The mean species abundance is obtained by noting that $M=Sm$. This is plotted in the main text.

\section{Phase diagram for heterogeneous interaction coefficients}

In this section, we set $\sigma_K=0$ so that $K_i=\mu_K$ for all $i\in\left\{{1,...,S}\right\}$. We seek to answer the following question: given a choice of $(\mu_c,\sigma_c)$ and a random draw of $c_{ij}$'s, what is the probability that the PA model is in the CR or the NR? 

\subsection{Boundary of the coexistence regime}
First, in the mean field approximation, we calculate the probability, $P_{\mathrm{CR}}$, that the system is in the CR.  We define the quantity
\be
\kappa_i =-\mu_K\mathrm{ln}\bigg\langle \exp\left(-\frac{1}{\mu_K}K_i^{\mathrm{eff}}(\vec{s})\right)\bigg\rangle,
\label{defkappa}
\ee
which is related the cumulants of the effective carrying capacity of species $i$. Recall that the CR occurs when $\kappa_i>0$ for all $i\in \left\{{1,...,S}\right\}$. Thus, to proceed, we need to explicitly calculate $\kappa_i$ in terms of $c_{ij}$. Based on numerical simulations, we see that the CR occurs only when $\mu_c<1$. By equation \ref{mfe}, $m\simeq 1$ when $\mu_c<1$, as long as $\omega\,\mathrm{ln}(\lambda)\ll\mu_K$, $\omega\ll \mu_K$. Therefore, $Q^\mathrm{CR}_i(s_i)\simeq \delta_{s_i,1}$, and we get 
\begin{align*}
	\bigg\langle \exp\left(-\frac{1}{\mu_K}K_i^{\mathrm{eff}}\right)\bigg\rangle_{Q^\mathrm{CR}}
		&= \bigg\langle \exp\left(-1+\sum_{j\not=i}c_{ij}s_j\right)\bigg\rangle_{Q^\mathrm{CR}}&\\
		&=e^{-1}\prod_{j\not=i}\langle e^{c_{ij}s_j}\rangle_{Q_i^\mathrm{CR}}&\\
		&=e^{-1}\prod_{j\not=i}e^{c_{ij}}&\\
		&=\exp\left(-1+\sum_{j\not=i}c_{ij}\right),& 
\end{align*}
which we can rewrite as
\begin{align*}
	\frac{1}{\mu_K}\kappa_i=1-\sum_{j\not=i}c_{ij}.
\end{align*}
This equation has a simple interpretation. Namely, the CR occurs when the effective carrying capacity of every species is positive in the presence of all species.
Thus, the probability that the ecosystem is in the CR, $P_{\mathrm{CR}}$, is just the probability that $\sum_{j\not=i}c_{ij}<1$. Since each $c_{ij}$ is drawn independently from a gamma distribution of mean $\mu_c/S$ and variance $\sigma_c^2/S$, it follows that $y_i:=\sum_{j\not=i}c_{ij}$ is gamma distributed with mean $\mu_c$ and variance $\sigma_c^2$ at leading order in large $S$; the explicit probability distribution is
\begin{align*}
	p_y(y_i)=\frac{1}{\theta_y^{k_y}\Gamma(k_y)}y_i^{k_y-1}\exp\left(-\frac{y_i}{\theta_y}\right),
\end{align*}
where $\Gamma$ denotes the Gamma function and
\begin{align*}
	k_y:=\frac{\mu_c^2}{\sigma_c^2},\;\;\;\; \theta_y:=\frac{\sigma_c^2}{\mu_c}.
\end{align*}
$P_{\mathrm{CR}}$ is the probability that $y_i<1$, i.e.
\begin{align*}
	P_{CR}=P(y_i<1)=\int_0^1\,\mathrm{d}y_i\,p_y(y_i)= \frac{1}{\Gamma(\mu_c^2/\sigma_c^2)}\gamma\left(\frac{\mu_c^2}{\sigma_c^2},\frac{\mu_c}{\sigma_c^2}\right),
\end{align*}
where $\Gamma$ is the gamma function and $\gamma$ is the lower incomplete gamma function. Given $(\mu_c,\sigma_c)$, this formula can be used to calculate $P_{CR}$ numerically, resulting in Fig. \ref{fig:analytics}.

\subsection{Boundary of the noisy regime}

Now, we find the probability that a random draw of $c_{ij}$'s causes the PA model to exhibit the NR. The NR occurs if $\kappa_i<0$ for all $i\in \left\{{1,...,S}\right\}$.  This can be easily calculated within the mean field approximation using (\ref{NRvar}). One gets
\begin{widetext}
	\begin{align*}
		\bigg\langle \exp\left(-\frac{1}{\mu_K}K_i^{\mathrm{eff}}\right)\bigg\rangle_{Q^\mathrm{NR}}=
			\int_{\mathbb{R}^S}\mathrm{d}\vec{s}\,Q^\mathrm{NR}(\vec{s})\exp\left(-\frac{1}{\mu_K}K_i^{\mathrm{eff}}\right)
			= e^{-1}\prod_{j\not=i}\frac{1}{\sqrt{2\pi\sigma_m^2}}
				\int_{-\infty}^{+\infty}\mathrm{d}s_j\,\exp\left( 
					-\frac{1}{2\sigma_m^2}[s_j-m]^2+c_{ij}s_j\right).
	\end{align*}
	Performing the Gaussian integrals yields
	\begin{align*}
		\bigg\langle \exp\left(-\frac{1}{\mu_K}K_i^{\mathrm{eff}}\right)\bigg\rangle_{Q^\mathrm{NR}}=\exp\left(-1+m\sum_{j\not=i}c_{ij}+\frac{1}{2}\sigma_m^2\sum_{j\not=i}c_{ij}^2 \right).
	\end{align*}
\end{widetext}
This can be rewritten as
\begin{equation} \label{kappac}
	\frac{1}{\mu_K}\kappa_i= 1-m\sum_{j\not=i}c_{ij}-\frac{1}{2}\sigma_m^2\sum_{j\not=i}c_{ij}^2.
\end{equation}
We further make the approximation
\begin{equation}\label{UncontrolledApproximation}
	\sum_{j\not=i}c_{ij}^2\simeq \mu_c\sum_{j\not=i}c_{ij},
\end{equation}
yielding the expression
\begin{align*}
	\frac{1}{\mu_K}\kappa_i\simeq 1-\left(m+\frac{1}{2}\mu_c\sigma_m^2\right)\sum_{j\not=i}c_{ij}
\end{align*}

It is useful to define a new random variable  $y_i:=\left(m+\frac{1}{2}\mu_c\sigma_m^2\right)\sum_{j\not=i}c_{ij}$. Recall that, in the NR, all $\kappa_i<0$, or equivalently that $y_i>1$. Thus, the probability of being in the noisy regime, $P_{\mathrm{NR}}$, is simply the probability that $y_i>1$. One can show that to leading order in $S$, $y_i$ is gamma distributed with mean $\left(m+\frac{1}{2}\mu_c\sigma_m^2\right)\mu_c$ and variance $\left(m+\frac{1}{2}\mu_c\sigma_m^2\right)\sigma_c^2$. Using this observation, one sees that 
\begin{widetext}
\begin{align*}
	P_{\mathrm{NR}}\simeq 1-P(y_i<1)= 1-\frac{1}{\Gamma\left(\frac{\mu_c^2}{\sigma_c^2}[m+\frac{1}{2}\mu_c\sigma_m^2]\right)}\gamma\left(\frac{\mu_c^2}{\sigma_c^2}[m+\frac{1}{2}\mu_c\sigma_m^2],\frac{\mu_c}{\sigma_c^2}\right),
\end{align*} 
\end{widetext}

\subsection{Alternative interpretation of the NR boundary}

As mentioned in the main text, $P_{\mathrm{NR}}$ has an alternative interpretation, namely the probability that $K_i^{\mathrm{eff}}(\vec{s})<0$ when at least $M+\Delta$ species are present, where $M:=\sum_{i=1}^{i=S}[\langle s_i\rangle]_{av}$ is the mean number of species and
\begin{align*}
	\Delta:=\frac{1}{\mu_c\mu_K}\sum_{i=1}^{i=S}[\delta K_i^{\mathrm{eff}}]_{av}
\end{align*}
is a threshold fluctuation in the number of species above the mean. 

We justify this interpretation now. First, $M$ is given by $Sm$, where $m$ satisfies equation \ref{mfe}. To compute $\Delta$, we note that $\delta K_i^{\mathrm{eff}}=\langle K_i^{\mathrm{eff}}\rangle-\kappa_i$. Given that $\langle K_i^{\mathrm{eff}}\rangle=\mu_K-\mu_Km\sum_{j\not=i}c_{ij}$, we use our approximate expression for $\kappa_i$ to obtain
\begin{align*}
	\delta K_i^{\mathrm{eff}}=\langle K_i^{\mathrm{eff}}\rangle-\kappa_i=\frac{1}{2}\sigma_m^2\mu_c\mu_K\sum_{j\not=i}c_{ij}.
\end{align*}
Thus, averaging over $c_{ij}$'s yields $[\delta K_i^{\mathrm{eff}}]_{av}=\frac{1}{2}\sigma_m^2\mu_c^2\mu_K$, from which we get
\begin{align*}
	\Delta=\frac{1}{2}S\mu_c\sigma_m^2.
\end{align*}
When $M+\Delta$ species are present, we have 
\begin{align*}
	K_i^{\mathrm{eff}}(\vec{s})=\mu_K-\mu_K\sum_{j\not= i}c_{ij}s_j=
		\mu_K-\mu_K\sum_{j=1}^{j=M+\Delta}c_{ij},
\end{align*}
where it is understood that $c_{ii}=0$. The quantity $\sum_{j=1}^{j=M+\Delta}c_{ij}$ is, to leading order in $S$, a gamma distributed random variable of mean $\frac{1}{S}(M+\Delta)\mu_c$ and variance $\frac{1}{S}(M+\Delta)\sigma_c^2$, a property shared by $\frac{1}{S}(M+\Delta)\sum_{j\not=i}c_{ij}$ at leading order. Thus, the probability that $K_i^{\mathrm{eff}}(\vec{s})<0$ is equal to the probability that $\mu_K-\mu_K\frac{1}{S}(M+\Delta)\sum_{j\not=i}c_{ij}<0$. Plugging in our expressions for $M$ and $\Delta$, we see that this is equal to the probability that
\begin{align*}
	\kappa_i\simeq\mu_K-\mu_K\left(m+\frac{1}{2}\mu_c\sigma_m^2\right)\sum_{j\not=i}c_{ij}<0,
\end{align*}
which is exactly $P_{\mathrm{NR}}$ obtained above.

\section{Phase diagram for heterogeneous carrying capacities}

Now set $\sigma_c=0$ so that $c_{ij}=\mu_c/S$ for all $i\in\left\{{1,...,S}\right\}$. We calculate the probabilities $P_{\mathrm{CR}}$ and $P_{\mathrm{NR}}$ that, given $(\mu_c,\sigma_K/\mu_K)$ and a random draw of $K_{i}$'s, PA model is in the CR or the NR.

\subsection{Boundary of the coexistence regime}

Recall, that $P_{\mathrm{CR}}$ is the probability that $\kappa_i>0$ for all $i\in \left\{{1,...,S}\right\}$. Since all species are present in the CR, our mean field variational ansatz (\ref{CRvar}) reduces to $Q_i^\mathrm{CR}(s_i)\simeq \delta_{s_i,1}$. Using this ansatz, one gets
\begin{align*}
	\bigg\langle \exp&\left(-\frac{1}{\mu_K}K_i^{\mathrm{eff}}\right)\bigg\rangle_{Q^\mathrm{CR}}&\\
		&= \bigg\langle \exp\left(-\frac{1}{\mu_K}K_i+\frac{\mu_c}{S\mu_K}\sum_{j\not=i}K_js_j\right)\bigg\rangle_{Q^\mathrm{CR}}&\\
		&=\exp\left(-\frac{1}{\mu_K}K_i+\frac{\mu_c}{S\mu_K}\sum_{j\not=i}K_j\right),& 
\end{align*}
which we can rewrite using (\ref{defkappa}) as
\begin{align*}
	\kappa_i=K_i-\frac{\mu_c}{S}\sum_{j\not=i}K_j.
\end{align*}
This is simply $K_i^{\mathrm{eff}}(\vec{s})$ when all species are present, and $P_{\mathrm{CR}}$ is the probability that this is positive. Since we draw each $K_i$ independently from a log-normal distribution, $p_K$, with mean $\mu_K$ and variance $\sigma^2_K$, we may apply the Central Limit Theorem to $\frac{1}{S}\sum_{j\not= i}K_j$ in the limit $S\rightarrow \infty$. In particular, $\frac{1}{S}\sum_{j\not= i}K_j$ approaches a Gaussian distribution of mean $\mu_K$ and variance $\sigma_K^2/S$, plus some terms of higher order in $1/S$. Thus, as $S\rightarrow \infty$,
\begin{align*}
	\frac{1}{S}\sum_{j\not= i}K_j\simeq\mu_K,
\end{align*}
which is to say that 
\begin{align*}
	\kappa_i\simeq K_i-\mu_c\mu_K
\end{align*}
when $S$ is very large. Therefore, $P_{\mathrm{CR}}$ is the probability that $K_i>\mu_c\mu_K$, or
\begin{widetext}
\begin{align*}
	P_{\mathrm{CR}}\simeq P(K_i>\mu_c\mu_K)=1-\int_0^{\mu_c\mu_K}\,\mathrm{d}K_i\,p_K(K_i)=1-\Phi\left(\frac{\log(\mu_c\mu_K)-l_K}{z_K}\right),
\end{align*}
where $\Phi$ is the normal cumulative distribution function and
\begin{align*}
	l_K:=\mathrm{ln}\left(\frac{\mu_K^2}{\sqrt{\mu_K^2+\sigma_K^2}}\right),\;\;\;\;
		z_K:=\sqrt{\mathrm{ln}\left(1+\frac{\sigma_K^2}{\mu_K^2}\right)}.
\end{align*}

\subsection{Boundary of the noisy regime}

We compute the probability $P_{\mathrm{NR}}$ that $\kappa_i<0$, using the Gaussian variational distribution (\ref{NRvar}). To do so, one  evaluates the appropriate Gaussian integrals:
	\begin{align*}
		\bigg\langle \exp\left(-\frac{1}{\mu_K}K_i^{\mathrm{eff}}\right)\bigg\rangle_{Q^\mathrm{NR}}&=
			\int_{\mathbb{R}^S}\mathrm{d}\vec{s}\,Q^\mathrm{NR}(\vec{s})\exp\left(-\frac{1}{\mu_K}K_i^{\mathrm{eff}}\right)\\
			&= e^{-K_i/\mu_K}\prod_{j\not=i}\frac{1}{\sqrt{2\pi\sigma_m^2}}
				\int_{-\infty}^{+\infty}\mathrm{d}s_j\,\exp\left( 
					-\frac{1}{2\sigma_m^2}[s_j-m]^2+\frac{\mu_c}{S\mu_K}K_js_j\right)&\\
			&=\exp\left(-\frac{1}{\mu_K}K_i+m\frac{\mu_c}{S\mu_K}\sum_{j\not=i}K_j+\frac{1}{2}\sigma_m^2\frac{\mu_c^2}{S^2\mu_K^2}\sum_{j\not=i}K_j^2 \right).&
	\end{align*}
Therefore
\begin{align*}
	\kappa_i= K_i-m\mu_c\frac{1}{S}\sum_{j\not=i}K_j-\frac{1}{2}\sigma_m^2\frac{\mu_c^2}{S\mu_K}\frac{1}{S}\sum_{j\not=i}K_j^2.
\end{align*}
For $S$ large, we can replace  $\frac{1}{S}\sum_{j\not=i}K_j$ and $\frac{1}{S}\sum_{j\not=i}K_j^2$ by the expectation value $K_i$ and $K_i^2$, respectively:
\begin{align*}
	\frac{1}{S}\sum_{j\not=i}K_j\simeq \mu_K,\;\;\; \frac{1}{S}\sum_{j\not=i}K_j^2\simeq\mu_K^2+\sigma_K^2.
\end{align*}
Substituting these expressions yields
\begin{align*}
	\kappa_i\simeq K_i-\mu_c\mu_K\bigg[m+\frac{1}{2S}\mu_c\left(1+\frac{\sigma_K^2}{\mu_K^2}\right)\sigma_m^2\bigg].
\end{align*}
For simplicity of notation, define $\delta:=\frac{1}{2S}\mu_c\left(1+\frac{\sigma_K^2}{\mu_K^2}\right)\sigma_m^2$. The probability that $\kappa_i<0$ is
\begin{align*}
	P_{NR}\simeq P(K_i<\mu_c\mu_K[m+\delta])=\int_0^{\mu_c\mu_K[m+\delta]}\,\mathrm{d}K_i\,p_K(K_i)=\Phi\left(\frac{\log(\mu_c\mu_K[m+\delta])-l_K}{z_K}\right),
\end{align*}
where $\Phi$, $l_K$, and $z_K$ are defined as above.
\end{widetext}
Once again $P_{\mathrm{NR}}$ has an interpretation as the probability that $K_i^{\mathrm{eff}}(\vec{s})<0$, when at at least $M+\Delta$ species are present, where $M:=\sum_{i=1}^{i=S}[\langle s_i\rangle]_{av}$ and $\Delta:=\frac{1}{\mu_c\mu_K}\sum_{i=1}^{i=S}[\delta K_i^{\mathrm{eff}}]_{av}$. Using a calculation analogous to the one presented in the previous section, it is straightforward to show that $\Delta=S\delta$.

\section{Computing $P_{\mathrm{NR}}$ directly from $Q$}

Here, we show that one can compute $\kappa_i$ for the NR using the mean-field $Q_i(s_i)=m\delta_{s_i,1}+(1-m)\delta_{s_i,0}$, without the Gaussian approximation to the variational distribution (\ref{NRvar}). We first restrict ourselves to the case where only the carrying capacities are heterogeneous ($\sigma_c=0$). One can obtain the same answer as above by taking the large $S$ limit. To see this, we expand $\kappa_i$ into powers of $\frac{1}{S}$. One has
\begin{widetext}
\begin{align*}
	\bigg\langle \exp\left(-\frac{1}{\mu_K}K_i^{\mathrm{eff}}\right)\bigg\rangle_Q
		&= \bigg\langle \exp\left(-\frac{1}{\mu_K}K_i+\frac{\mu_c}{S\mu_K}\sum_{j\not=i}K_js_j\right)\bigg\rangle_Q&\\
		&=e^{-K_i/\mu_K}\prod_{j\not=i}\bigg\langle  \exp\left(\frac{\mu_c}{S\mu_K}K_js_j\right)\bigg\rangle_Q&\\
		&=e^{-K_i/\mu_K}\prod_{j\not=i}\bigg[ m \exp\left(\frac{\mu_c}{S\mu_K}K_j\right)+(1-m)\bigg]&\\
		&=\exp\left(-\frac{K_i}{\mu_K}+\sum_{j\not=i}\mathrm{ln}\bigg[1+ m \exp\left(\frac{\mu_c}{S\mu_K}K_j\right)-m\bigg]\right),&
\end{align*}
yielding
\begin{align*}
	\kappa_i=K_i-\mu_K\sum_{j\not=i}\mathrm{ln}\bigg[1+ m \exp\left(\frac{\mu_c}{S\mu_K}K_j\right)-m\bigg]
			=K_i-\mu_K\sum_{j\not=i}\sum_{n=1}^{\infty}\frac{1}{n}(-1)^{n+1}m^n\bigg[\sum_{l=1}^{\infty}\frac{1}{l!}\left(\frac{\mu_cK_j}{S\mu_K}\right)^l\bigg]^n,
\end{align*}
where we employed the series expansions of both $\mathrm{ln}(1+x)$ and $\exp(x)-1$. Now, neglecting terms of third order or higher in $\frac{1}{S}$, we get
\begin{align*}
	\kappa_i\simeq K_i-\mu_K\sum_{j\not=i}\sum_{n=1}^{\infty}\frac{1}{n}(-1)^{n+1}m^n\bigg[\frac{\mu_cK_j}{S}+\frac{1}{2}\left(\frac{\mu_cK_j}{S\mu_K}\right)^2\bigg]^n\simeq 
		K_i-\mu_K\sum_{j\not=i}\bigg[m\frac{\mu_cK_j}{S\mu_K}+m(1-m)\frac{1}{2}\left(\frac{\mu_cK_j}{S\mu_K}\right)^2\bigg].
\end{align*}
Using the identity $\sigma_m^2=m(1-m)$, we obtain 
\begin{align*}
	\kappa_i\simeq K_i-m\mu_c\frac{1}{S}\sum_{j\not=i}K_j-\frac{1}{2}\sigma_m^2\frac{\mu_c^2}{S\mu_K}\frac{1}{S}\sum_{j\not=i}K_j^2,
\end{align*}
which is exactly what we obtained using (\ref{NRvar}).
\end{widetext}

A similar expansion can be employed in the case of heterogeneous $c_{ij}$, but the limit is more delicate. Higher order terms cannot be neglected as $S\rightarrow\infty$ because, in this limit, the moment $\sum_{j\not=i}c_{ij}^n$ does not vanish for any positive integer $n$. However, these higher order moments can be neglected in the limit $\mu_c\gg \sigma_c^2$ (provided that 
$S\gg \mu_c^2/\sigma_c^2$), which, according to our numerical simulations, is consistent with the NR. This follows from the form of the moment-generating function of gamma distribution:
\begin{align*}
	M(x)=\exp\bigg[\frac{1}{S}\frac{\mu_c^2}{\sigma_c^2}\mathrm{ln}\left(1-\frac{\sigma_c^2}{\mu_c}x\right)\bigg].
\end{align*}

\bibliography{refs_ecology}

\end{document}